\documentclass[10pt]{article}
\usepackage{psfrag,epsf}
\usepackage{url} % not crucial - just used below for the URL 

\newcommand{\blind}{0}

\usepackage{natbib}

\usepackage{bbm}
\usepackage[a4paper, top=3cm, bottom=3cm, left=3cm, right=3cm]{geometry}
\usepackage[plain]{fullpage}
\usepackage[normalem]{ulem}
\usepackage[latin1]{inputenc}
\usepackage{comment}
\usepackage{graphics}
\usepackage{tabularx}
\usepackage{array,multirow}

\usepackage{setspace,color}
\usepackage{graphicx,color}
\usepackage{amsmath, amsfonts, amssymb, amstext,enumerate}
\usepackage{mathptmx}
\usepackage{times,psfrag,graphicx,theorem,epsfig,lscape,amssymb,amsmath,dcolumn,
multirow,epic,float,enumerate,latexsym,ifthen,pifont,comment,bm,subfigure,psfrag,epsf}
\usepackage{dsfont}
\usepackage{xr}

\theoremstyle{plain} \newtheorem{assumption}{Assumption}
\theoremstyle{plain} 
\theoremstyle{plain} \newtheorem{proposition}{Proposition}
\theoremstyle{plain}

\newcommand{\bW}{\bm{W}}
\newcommand{\bw}{\bm{w}}

\newcommand{\bY}{\bm{Y}}

\newcommand{\bX}{\bm{X}}

\newcommand{\balpha}{\bm{\alpha}}

\newcommand{\bbeta}{\bm{\beta}}
\newcommand{\bgamma}{\bm{\gamma}}

\newcommand{\btheta}{\bm{\theta}}

\newcommand{\bsigma}{\bm{\sigma}}

\begin{document}

\def\spacingset#1{\renewcommand{\baselinestretch}%
{#1}\small\normalsize} \spacingset{1}

%%%%%%%%%%%%%%%%%%%%%%%%%%%%%%%%%%%%%%%%%%%%%%%%%%%%%%%%%%%%%%%%%%%%%%%%%%%%%%

\if0\blind
{
  \title{\bf Bayesian Inference for Sequential Treatments under Latent Sequential Ignorability}
  \author{Federico Ricciardi\textsuperscript{1},
  	Alessandra Mattei\textsuperscript{2}, and Fabrizia Mealli\textsuperscript{2} %\thanks{The authors gratefully acknowledge \textit{please remember to list all relevant funding sources in the unblinded version}}
  \hspace{.2cm}\\
  \textsuperscript{1}Department of Statistical Science, University College London, UK\\
    \textsuperscript{2}Department of Statistics, Computer Science, Applications, University of Florence, Italy   
      }
  \maketitle
} \fi

\if1\blind
{
  \bigskip
  \bigskip
  \bigskip
  \begin{center}
    {\LARGE\bf Bayesian Inference for Sequential Treatments under Latent Sequential Ignorability}
\end{center}
  \medskip
} \fi

\bigskip

\begin{abstract}
We focus on causal inference for longitudinal treatments, where units are assigned to treatments at multiple time points, aiming to assess the effect of different treatment sequences on an outcome observed at a final point. A common assumption in similar studies is Sequential Ignorability (SI): treatment assignment at each time point is assumed independent of
future potential outcomes given past observed outcomes and covariates. SI is questionable when treatment participation depends on individual choices, and treatment assignment may depend on unobservable quantities associated with future outcomes. We rely on Principal Stratification to formulate a relaxed version of SI: Latent Sequential Ignorability (LSI) assumes that treatment assignment is conditionally independent on future potential outcomes given past treatments, covariates and principal stratum membership, a latent variable defined by the joint value of observed and missing intermediate outcomes. We evaluate SI and LSI, using theoretical arguments and simulation studies to investigate the performance of the two assumptions when one holds and inference is conducted under both. Simulations show that when SI does not hold, inference performed under SI leads to misleading conclusions. Conversely, LSI generally leads to correct posterior distributions, irrespective of which assumption holds.
\end{abstract}

%\keywords{Latent sequential ignorablity, Longitudinal treatments, Principal stratification, Sequential ignorablity, Rubin Causal Model.}

\noindent%
\vfill

\newpage
\spacingset{1.45} % DON'T change the spacing!

\section{Introduction}
\label{sec:Intro}
Many observational studies in different fields, including economics, social science and epidemiology, are often interested in the evaluation of causal effects of time-varying treatments, which are assigned to units sequentially over time \cite[e.g.,][]{Robins:1986, Robins:1989,  Robins:1997, RHB:2000, GillRobins:2001,  Lechner:2009, AchyBrou_et_al:2010, Zajonc:2012, ImaiRatkovic2014, ZEL:2018}.

In the presence of time-varying treatments, causal inference is challenging  because intermediate variables are simultaneously post-treatment outcomes and pretreatment confounders. Therefore the analysis of time-varying treatments requires  methodological tools that can properly account for a growing number of intermediate variables, some of which are only partially observed, and sequential selection. In this paper we propose to face these challenges when  assessing the effect of different sequences of a time-varying treatment on some final outcome observed at the end of the study.

We will frame our discussion in the context of the potential outcomes approach to causal inference, also referred to as the Rubin Causal Model \cite[RCM, e.g.,][]{Rubin:1974, Rubin:1977, Rubin:1978, Holland:1986}.
A critical part of the RCM is the formulation of a treatment assignment mechanism, and this task is even more crucial in longitudinal studies. An assumption usually invoked in evaluation studies with longitudinal treatment is \textit{Sequential Ignorability} \cite[SI,][]{Robins:1986}, which amount to assuming that the observed treatment at a given time point is independent of future potential outcomes given past observed outcomes, past treatments and covariates up to that point. 
Sequential ignorability may be a reasonable assumption in various settings. For instance, in medicine, physicians may propose therapies randomly conditional on observed patient's characteristic, prognostic factors and prior treatments up to that point. In labor economics caseworkers may randomly offer training programs to participants conditional on previous training program participation up to that point and  observed performances.

On the other hand, and especially for observational studies or in settings where participation in the treatment depends on individual choices, treatment assignment may depend on unobservable quantities associated with future potential outcomes as well as on unobserved past potential outcomes, even conditional on the observed history, so that  the sequential ignorability assumption fails to hold. 
For instance, in program evaluation, subjects may decide to participate in a program at a given time point using both information on their performances under the treatments previously received (the observed outcomes), which also the experimenter can observe, as well as information on their performances under alternative unobserved treatment sequences (the missing outcomes), which may be known to subjects (maybe with some approximation) but unknown to the experimenter. In medicine, the treatment a patient decides to take at a given time point may depend on both the observed patient's history (including previous treatments and observed outcomes)  as well as on some unobserved patient's characteristic related to the missing outcomes.

In order to relax SI, we rely on Principal Stratification \cite[PS,][]{FrangakisRubin:2002} and we formulate a milder version of SI that we call \textit{Latent Sequential Ignorability} (LSI). LSI assumes that treatment assignment is conditionally independent on future potential outcomes given pre-treatment variables, past treatments, and principal strata, defined by the joint value of observed and missing intermediate outcomes up to that point. 
Principal strata encode personal characteristics reflected in the intermediate outcomes, therefore if intermediate outcomes are associated with future treatment and  outcomes they can be viewed as a coarsened representation of the latent unobserved structure that may affect the decision to participate in the treatment \citep{MealliMattei:2012}.
Alternative assumptions could be considered, e.g., by gleaning from the literature on non-ignorable missing data, but we look at LSI as a valuable starting point to move forward the traditional SI assumption.

LSI has appealing features, but also raises challenging inferential issues due to the latent nature of principal strata. We propose the Bayesian approach for inference, which is particularly useful for accounting for uncertainties and for pooling information from the data in complex settings. Under SI causal estimands, such as average causal effects,  are usually point identified, that is, they can be expressed as known function of the distribution of the observed data. Under LSI, some parameters may be partially   identified in the sense 
 that multiple values of the
 	parameter can correspond to the same distribution of observables, and thus the parameters
 	cannot be consistently estimated, but the possible set of values for the parameters which
 	are consistent with the observed data law is smaller than the a-priori set of possible values: the observed data law gives an identification region for partially identified parameters  \cite[e.g.,][]{Gustafson:2010}.
The Bayesian approach is particularly appealing to draw inference on  partially  identified parameters. In fact, Bayesian inference is based on the posterior distribution of the parameters of interest, which are derived by updating a prior distribution to a posterior distribution via a likelihood, irrespective of whether the parameters are fully or partially identified, and if the prior is proper, the posterior distribution will be proper, too. 
Bayesian analysis conducted under LSI naturally provides a framework for  sensitivity analysis with respect to specific violations of SI, where sensitivity parameters are meaningful quantities, with a direct interpretation
(see Section~\ref{sec:PrW2} for details).

In this work we discuss and compare  sequential ignorability and latent sequential ignorability, using both theoretical arguments and simulation studies in which we investigate the relative performance of the two alternative assumptions when, in turn, one holds and inference is conducted under both assumptions. We also illustrate our framework using real data on financial aids to firms to investigate the effectiveness of interests free loans on firms' employment policies. 
In this study firms may have access to public loans multiple times over subsequent years and our focus is on contrasting firms' performances measured in terms of employment levels at the end of the study under different treatment sequences \cite[][]{PiraniMarianiMealli2013}. 
Our observation period is from 2002 to 2007 and we focus on assessing causal effects on the number of employees of receiving at least a loan either between 2002 and 2004 or between 2005 and 2007. 
The hiring policy of a firm in 2004 is an intermediate post-treatment variable that may be reasonably associated with both the number of employees in 2007 as well as the decision to apply for a loan between 2005 and 2007.   
Sequential ignorability implies that the probability of receiving at least a loan between 2005 and 2007 does not depend on
the potential outcomes for the number of employees in 2007 conditional on covariates, loan status in 2004, and the observed firm's  hiring policy in 2004. We argue that this assumption may be debatable, 
but the latent sequential ignorability may be more reasonable because the decision of a firm to apply for a loan between 2005 and 2007 may reasonably depend on both the observed firm's hiring policy in 2004 and the unobserved firm's  hiring policy in 2004 that the firm would have adopted under the alternative loan status in 2004.

Throughout the article we focus on assessing causal effects of a specified longitudinal treatment on an  outcome that would have been observed at the end of the study. A valuable topic for future research is the extension of our framework to the evaluation of  dynamic treatment regimes, which usually describe  adaptive policies that propose actions in each treatment period depending on past observations and decisions \cite[e.g.,][]{HeckmanNavarro:2007, HongRaudenbush:2008, Murphy:2003, Robins:2004, Zajonc:2012}.

The article is organized as follows. In Section~\ref{sec:Basic} we introduce the framework and the causal estimands we focus on. In Section~\ref{sec:Assumptions} we formally define the assignment mechanism and the critical assumptions, SI and LSI. 
In Section~\ref{sec:PrW2} we compare  SI and LSI,  highlighting their implications and showing how latent sequential ignorability provides a natural framework for assessing the robustness of the estimates to  specific violations of the sequential ignorability assumption.
In Section \ref{sec:Inference} we discuss the inferential challenges arising with longitudinal treatments, briefly reviewing the existing approaches to address them, which are mainly based on SI. We then describe the Bayesian framework for inference, a natural and appealing approach that also allows us to make comparisons between SI and LSI on the same ground.
In Section \ref{sec:Simulations} we investigate  the role and implications of the two alternative assumptions using some simulated experiments. In Section \ref{sec:Case} we conduct causal inference under SI and LSI in the context of the illustrative case study. Finally, we conclude with a discussion in Section \ref{sec:conclusion}.

\section{Basic Setup}
\label{sec:Basic}

In this article we will focus on a simple setup with a two-period structure and binary treatments.
This simplified setting allows us to clearly describe all the conceptual issues surrounding sequential treatments, avoiding technical complications that may mask our primary objective, that is, highlighting the implications of SI and LSI and comparing inferences under the two assumptions.
Indeed, the extension to more time points makes notation more complicated, but does not represent an issue for the theoretical framework, although it may raise inferential and computational challenges. 

\subsection{Notation}
\label{sec:Notation}
Consider a group of units  indexed by $i \in \{1, \dots, n\}$.
In each of two periods, indexed by $t=1,2$, units 
can be potentially assigned either an active treatment ($w_t = 1$) or a control treatment, which may be no treatment at all ($w_t = 0$).
Let $W_{it}$ denote the treatment unit $i$ actually receives at time $t$: $W_{it}=1$  if the unit
is exposed to the active treatment, $W_{it}=0$ if the unit is exposed to the control treatment.
Let $\bW_i = \left(W_{i1}, W_{i2}\right)$. Then $\bW_i  \in \left\{(0,0),(1,0), (0,1), (1,1)  \right\}$, that is, units can  experience treatment in  neither period, $\bW_i  = (0,0)$; only in the first period, $\bW_i  = (1,0)$;  only in the second period, $\bW_i  = (0,1)$; or in both periods, $\bW_i  = (1,1)$. Let $\bW_{t}$ denote the $n$-dimensional vector with $i$-th element $W_{it}$, which is a random vector prior to the assignment at time $t$, and let $\bw_{t}$ be a realization of the random vector $\bW_{t}$.  

Let $Y_{i2}$ denote the final outcome, which is the object of primary interest and it is measured after  assignment of the final treatment, $\bW_2$.  After assignment to the first treatment, but prior to the assignment to the second treatment,  an intermediate outcome, $Y_{i1}$, can be measured for each unit $i$.
The intermediate variable we consider is the lagged outcome (or a transformation of the lagged outcome), which is a measure of the same substantive quantity as the final outcome, but measured at a previous time-point between  the receipt of the first treatment and  the receipt of the final treatment.   This choice is compelling, since it is reasonable to believe that the lagged intermediate outcome is related to both the treatment assignment at time $t=2$, $\bW_2$, and the final outcome, $Y_{i2}$. For instance in our illustrative study, units are firms and $W_{i1}$ is equal to $1$ if firm $i$ receives a loan in the first period (2002 to 2004); and, similarly,  $W_{i2}$  is equal to 1 if firm $i$ receives a loan in the second period  (2005 to 2007). The  final outcome of interest is   the number of employees at the end of 2007. As intermediate outcome   we consider a binary indicator equal to 1 for firms that increase their number of employees by the end of 2004 and zero otherwise. The choice of a firm to hire new employees during the first treatment period is reasonably related to the the decision to apply for a loan in the second treatment period as well as the number of employees in 2007.

For each unit $i$, let $Y_{i1}(\bw_1)$ denote the potential outcomes for the intermediate variable at time $t=1$ given treatment assignment $\bw_1$ in the first period, and let $Y_{i2}(\bw_1, \bw_2)$ denote the  potential outcome  for the final outcome given the entire treatment assignment sequence, $(\bw_1, \bw_2)$. 
 
We make the Stable Unit Treatment Value Assumption \cite[SUTVA,][]{Rubin:1980}, stating that potential outcomes for any unit are unaffected by the treatment assignments of other units (no interference), and that for each unit there are no different versions of treatment. Formally, 
\begin{assumption}\label{ass:SUTVA} SUTVA
\begin{flushleft}
\begin{tabular}{l}
 If $w_{i1}=w_{i1}'$, then $Y_{i1}(\bw_1)=Y_{i1}(\bw_1')$; \\
 If $(w_{i1}, w_{i2})=(w'_{i1}, w'_{i2})$, then $Y_{i2}(\bw_{1}, \bw_{2})=Y_{i2}(\bw'_{1}, \bw'_{2})$.
\end{tabular}
\end{flushleft}
\end{assumption}
SUTVA allows us to write $Y_{i1}(\bw_1)=Y_{i1}(w_{i1})$ and  $Y_{i2}(\bw_1, \bw_2)=Y_{i2}(w_{i1}, w_{i2})$, therefore for each unit $i$  there are two potential outcomes for the post-treatment intermediate variable measured after assignment to the first treatment, $Y_{i1}(0)$ and $Y_{i1}(1)$,  and four potential outcomes for the final outcome, $Y_{i2}(0,0)$, $Y_{i2}(1,0)$, $Y_{i2}(0,1)$ and $Y_{i2}(1,1)$.

\subsection{Causal Estimands}
\label{sec:CE}
Causal effects on the final outcome, $Y_{i2}$, are defined at the unit-level as comparisons of potential outcomes for the final outcome under alternative treatment sequences.
For instance, a causal effect of the treatment sequence $(w_1, w_2)$ versus the treatment sequence 
$(w'_1, w'_2)$ for a unit $i$ is defined as a comparison between the potential outcomes $Y_{i2}(w_1, w_2)$  and  $Y_{i2}(w_1', w_2')$. Estimands of interest may be simple differences $Y_{i2}(w_1, w_2)-Y_{i2}(w_1', w_2')$,  but in general comparisons can take different forms.
Causal effects can also be defined for collections of units. More generally, causal effects are comparisons between potential outcomes for a common set of units \cite[][]{FrangakisRubin:2002, Rubin:2005}. 
In this article we consider the $n$ units as a random sample from a large superpopulation, and we focus  on population Average Treatment Effects (ATEs) on the final outcome, that is, the expected value of the difference between potential outcomes at time $t=2$ under different treatment sequences.
In the presence of two-period binary treatments, we have:
\begin{equation}
\label{eq:ATE}
ATE_{w_1w_2.w_1'w_2'} = E[Y_{i2}(w_1, w_2) - Y_{i2}(w_1', w_2')], \qquad \hbox{for } (w_1, w_2) \neq (w_1', w_2') \in \{0,1\}^{2}.
\end{equation}
We focus on six causal effects by comparing the following treatment sequences: 
$(1,1)$ versus $(1,0), (0,1)$ and $(0,0)$;
$(1,0)$ versus $(0,1)$ and $(0,0)$; and $(0,1)$ versus $(0,0)$. For instance, in our illustrative example based on real data,   $ATE_{11.00}$ is the difference between the  average number of employees if all firms received at least a loan in both treatment periods  and the  average number of employees if all firms received a loan in neither treatment period.

\section{The Assignment Mechanism}
\label{sec:Assumptions}

The fundamental problem of causal inference \citep{Holland:1986, Rubin:1978} is that for each unit we can only observe at most one of the potential outcomes for each post-treatment variable.
In our setting with two-period binary treatments, for each unit $i$ we observe one out of two intermediate potential outcomes at time $t=1$, i.e., $Y_{i1}^{obs}=Y_{i1}(W_{i1})$; and one out of four potential outcomes at time $t=2$, i.e. $Y_{i2}^{obs}=Y_{i2}(W_{i1}, W_{i2})$.
Potential outcomes under unassigned treatment sequences are missing: $Y_{i1}^{mis}=Y_{i1}(1-W_{i1})$ and
$\bY_{i2}^{mis}=\{Y_{i2}(1-W_{i1}, W_{i2}), Y_{i2}(W_{i1}, 1-W_{i2}); Y_{i2}(1-W_{i1}, 1-W_{i2})\}$.
Therefore, inference on causal effects requires to solve a missing data problem, which is particularly challenging in the presence of longitudinal treatments, even in the case with two-period binary treatments.

In order to learn about the causal effects of interest it is crucial to posit a treatment assignment mechanism. The assignment mechanism is a row-exchangeable function  of all covariates and of all potential outcomes, giving  the probability of any vector of treatment sequences. For each unit $i$, let  $\bX_i$ denote an observed vector of pre-treatment variables, variables that are not affected by treatments assignment.   The assignment mechanism for a two-period treatment can be formally defined as follows:

$$
Pr\left(\bW \mid \bX, \bY_{1}(0), \bY_{1}(1), \bY_{2}(0,0), \bY_{2}(1,0), \bY_{2}(0,1), \bY_{2}(1,1)\right)
$$
where $\bW$ is a $n\times 2$ matrix with $i$-th row equal to $\bW_i=(W_{i1}, W_{i2})$, $\bX$ is a matrix with $n$ rows and $i$-th row equal to $\bX_i$, and $\bY_{1}(w_1)$ and $\bY_{2}(w_1,w_2)$ are  $n-$dimensional vectors with $i$-th elements equal to
$Y_{i1}(w_1)$ and $Y_{i2}(w_1,w_2)$, respectively, for $w_1 \in \{0,1\}$ and $w_2 \in \{0,1\}$.

In longitudinal settings  the assignment mechanism is very complex. We consider two basic restrictions on the assignment mechanism, assuming that it is individualistic and probabilistic. Let
\begin{eqnarray*}
	\lefteqn{
p_i\left(\bw \mid \bX, \bY_{1}(0), \bY_{1}(1), \bY_{2}(0,0), \bY_{2}(1,0), \bY_{2}(0,1), \bY_{2}(1,1)\right) =}\\&& \sum_{\bW: \bW_i=\bw} Pr\left(\bW \mid \bX, \bY_{1}(0), \bY_{1}(1), \bY_{2}(0,0), \bY_{2}(1,0), \bY_{2}(0,1), \bY_{2}(1,1)\right)
\end{eqnarray*}
denote the unit-level assignment probabilities for $\bw \in \{0,1\}^2$. An assignment mechanism is 
individualistic if 
\begin{eqnarray*}
	\lefteqn{
		p_i\left(\bw \mid \bX, \bY_{1}(0), \bY_{1}(1), \bY_{2}(0,0), \bY_{2}(1,0), \bY_{2}(0,1), \bY_{2}(1,1)\right) =}\\&&  
Pr\left(\bW_{i} = \bw \mid \bX_i, Y_{i1}(0), Y_{i1}(1), Y_{i2}(0,0), Y_{i2}(1,0), Y_{i2}(0,1), Y_{i2}(1,1)\right) 
\end{eqnarray*}
for all $i=1,\ldots, n$ and  $\bw \in \{0,1\}^2$, and
\begin{eqnarray*}
	\lefteqn{
		Pr\left(\bW \mid \bX, \bY_{1}(0), \bY_{1}(1), \bY_{2}(0,0), \bY_{2}(1,0), \bY_{2}(0,1), \bY_{2}(1,1)\right) \propto}\\&&
		\prod_{i} \prod_{\bw \in \{0,1\}^2} 	p_i\left( \bw \mid \bX_i, Y_{i1}(0), Y_{i1}(1), Y_{i2}(0,0), Y_{i2}(1,0), Y_{i2}(0,1), Y_{i2}(1,1)\right)^{\mathbf{1}\{\bW_i=\bw\}}
%	\prod_{i} \prod_{\bw \in \{0,1\}^2} 	Pr\left(\bW_{i} = \bw \mid \bX_i, Y_{i1}(0), Y_{i1}(1), Y_{i2}(0,0), Y_{i2}(1,0), Y_{i2}(0,1), Y_{i2}(1,1)\right)^{\mathbf{1}\{\bW_i=\bw\}}
\end{eqnarray*}
for $\left(\bW, \bX, \bY_{1}(0), \bY_{1}(1), \bY_{2}(0,0), \bY_{2}(1,0), \bY_{2}(0,1), \bY_{2}(1,1)\right) \in \mathbb{A}$, for some set $\mathbb{A}$, and zero otherwise.

An assignment mechanism is  probabilistic if
$$
0< p_i\left(\bw\mid \bX_i, Y_{i1}(0), Y_{i1}(1), Y_{i2}(0,0), Y_{i2}(1,0), Y_{i2}(0,1), Y_{i2}(1,1)\right)<1, 
$$
for all $i=1,\ldots, n$, and $\bw \in \{0,1\}^2$.

Even under these restrictions, the assignment mechanism still remains complex, because it depends on a large number of missing values, $Y_{i1}^{mis}$ and $\bY^{mis}_{i2}$, for all $i$.
In order to reduce the complexity of the assignment mechanism, we now  
formulate some assumptions, which allow us to characterize longitudinal observational studies and draw inference on the causal estimands of interest. To this end, it is useful to factorize the unit-level assignment probabilities as product of  the assignment probabilities at time $t=1$ and the conditional assignment probabilities at time $t=2$ given the treatment received at time one. Formally, by the law of total probability, we have
\begin{eqnarray*}
\lefteqn{
Pr\left(\bW_{i} \mid \bX_i, Y_{i1}(0), Y_{i1}(1), Y_{i2}(0,0), Y_{i2}(1,0), Y_{i2}(0,1), Y_{i2}(1,1)\right)=}\\&&
Pr\left(W_{i1} \mid \bX_i, Y_{i1}(0), Y_{i1}(1), Y_{i2}(0,0), Y_{i2}(1,0), Y_{i2}(0,1), Y_{i2}(1,1)\right) \times \\&&
Pr\left(W_{i2}  \mid W_{i1}, \bX_i, Y_{i1}(0), Y_{i1}(1), Y_{i2}(0,0), Y_{i2}(1,0), Y_{i2}(0,1), Y_{i2}(1,1)\right).
\end{eqnarray*}

Much of the literature on time-varying treatments copes with the complications arising in the presence of sequential treatments by assuming that the assignment mechanism is sequentially ignorable \cite[][]{Robins:1986}:
\begin{assumption}\label{ass:SI}  Sequential Ignorability (SI)
	\begin{eqnarray}
		Pr\left(W_{i1}\mid \bX_i, Y_{i1}(0), Y_{i1}(1), Y_{i2}(0,0), Y_{i2}(1,0), Y_{i2}(0,1), Y_{i2}(1,1)\right)&=&Pr\left(W_{i1} \mid \bX_i\right) \label{eq:SI1} \\
Pr\left(W_{i2} \mid \bX_i,W_{i1}, Y_{i1}^{obs},  Y_{i2}(0,0), Y_{i2}(1,0), Y_{i2}(0,1), Y_{i2}(1,1) \right) &=& Pr\left(W_{i2} \mid \bX_i, W_{i1}, Y_{i1}^{obs}\right)   \notag %\label{eq:SI2} 
	\end{eqnarray}
\end{assumption}
SI implies that treatment assignment at each time point is independent of  all future potential outcomes given past observed outcomes, treatments and covariates. 

SI guarantees that, within cells defined by the pre-treatment covariates, the mean of the potential outcomes under a specific treatment sequence can be estimated from the observed data  as weighted average of the means of the observed final outcome under that treatment sequence across groups defined by the observed intermediate outcome, with weights that depend on the  distribution of the observed intermediate outcome. Formally, under SI
\begin{eqnarray*}
E\left[Y_{i2}(w_1, w_2) \mid \bX_i \right] = 
\int
E\left[Y^{obs}_{i2} \mid W_{i1}=w_1,W_{i2}=w_2,Y^{obs}_{i1}=y_1,  \bX_i\right] d\, F_{Y^{obs}_{i1}\mid W_{i1}=w_1, \bX_i}(y_1),
\end{eqnarray*}
where $F_{Y^{obs}_{i1}\mid W_{i1}=w_1, \bX_i}(\cdot)$ is the conditional cumulative distribution function of the intermediate outcome, $Y_1^{obs}$,
given the observed treatment at time $t=1$ and pre-treatment covariates. 

It is worth noting that SI defines the assignment mechanism at each time point separately and independently of the other time points. The underlying idea is that at each time point a new study has been conducted, for which an assignment mechanism must be posited, and  SI implies that at every time the treatment is as if randomized with probabilities depending on the observed history.
Although SI allows one to easily identify and estimate the conditional expectation of the potential outcomes of interest, it does not permit to reconstruct the assignment mechanism underlying the longitudinal study in its entirety, that is, the joint conditional probability 
of $\bW_{i}$ given all the potential outcomes and covariates. To this end we can introduce a different  ignorability  assumption, which is highly related to SI:
\begin{assumption}\label{ass:SI_AM} Sequential Ignorability of Longitudinal Treatment Assignment  (SIL)
\begin{eqnarray*}
	&Pr\left(\bW_{i}\mid \bX_i, Y_{i1}(0), Y_{i1}(1), Y_{i2}(0,0), Y_{i2}(1,0), Y_{i2}(0,1), Y_{i2}(1,1)\right)=&\\
	&Pr\left(W_{i1} \mid \bX_i\right) \times 
	Pr\left(W_{i2} \mid W_{i1}, Y_{i1}^{obs},  \bX_i\right).&
\end{eqnarray*}
\end{assumption}
Assumption  \ref{ass:SI_AM} amounts to assuming that treatment assignment at each time point is independent of  past missing potential outcomes and all future potential outcomes given past observed outcomes, treatments and covariates. 
Assumption  \ref{ass:SI_AM} is slightly stronger than   Assumption  \ref{ass:SI}, because it implies Assumption  \ref{ass:SI} but the converse is not true: Assumption  \ref{ass:SI} ignores the relationship between  treatment assignment at time $t=2$ and past missing potential outcomes, only requiring that   the assignment mechanism at time $t=2$ is independent of   all future potential outcomes conditional on the observed history. 
Nevertheless Assumptions  \ref{ass:SI} and \ref{ass:SI_AM}  have the same implications from an inferential perspective. For this reason, although Assumption  \ref{ass:SI}   is weaker than 
Assumption  \ref{ass:SI_AM}, in
practice it is difficult that a convincing argument can be made for the weaker Assumption  \ref{ass:SI}  without the argument being equally cogent for the stronger Assumption  \ref{ass:SI_AM}.

Sequential ignorability assumptions may be reasonable  in various settings, including longitudinal observational studies where it is reasonable to believe that treatments are sequentially assigned using only the observed information \cite[e.g.,][]{Zajonc:2012}. However, as in single point observational studies, where the usually made strong ignorability assumption may fail to hold due to the presence of  unobserved confounders associated  with both the potential outcomes and the treatment indicator \cite[][]{RosenbaumRubin:1983, Rosenbaum:1987, Imbens:2003, NIM:2008}, here  sequential ignorability may be arguable  due to the presence of time-varying unobserved confounder factors. 
The key insight is that the joint potential values of the intermediate outcome at time $t=1$, $(Y_{i1}(0),Y_{i1}(1))$, may represent an accurate summary of the
unobserved variables related to both treatment assignment  at time $t=2$ and the final outcome, due to which sequential ignorability assumptions do not hold.

Motivated by this intuition,  we use the concept of principal stratification \citep{FrangakisRubin:2002} to define  a new assumption on the longitudinal assignment mechanism, which may be a valuable alternative to sequential ignorability assumptions when they are assumed to fail in some specific and meaningful ways.
The joint potential values of the intermediate outcome at time $t=1$, $(Y_{i1}(0),Y_{i1}(1))$, defines a classification of units into principal strata. 
 Although the literature has mainly concentrated on studies with binary post-treatment variables,
principal stratification \textit{per se} does not require that the intermediate outcome is binary. Recent work has indeed considered the
application of principal stratification in the presence of  multi-valued categorical or continuous post-treatment variables
\cite[e.g.,][]{MatteiMealli:2007, JinRubin2008, SchwartzLiMealli:2011, frumento2012, feller2016}. 
Nevertheless the presence of categorical or continuous intermediate variables introduce serious challenges to principal stratification analysis.  
The number of principal strata  increases with the number of values in the support of the intermediate variable, and a continuous intermediate variable  generates a continuum of principal strata, leading to substantial complications in both inference and interpretation.
In the literature, flexible parametric \cite[e.g.,][]{JinRubin2008} 
and semi-parametric models \cite[e.g.,][]{SchwartzLiMealli:2011}, possibly
coupled with structural assumptions \cite[e.g.,][]{MatteiMealli:2007, JinRubin2008, feller2016}, have been developed to face identification and estimation issues arising
with a high (possibly uncountable) number of principal strata. 
In order to avoid additional complications  and complex model structures, which may mask our primary objectives, hereafter we prefer to consider a binary intermediate variable.
The (basic) principal stratification with respect to the binary intermediate outcome $Y_1$ classifies  units  into four groups according to the joint potential values of $Y_1$,   $Y_{i1}(0)$ and $Y_{i1}(1)$: 
$00 = \{i: (Y_{i1}(0)=0, Y_{i1}(1)=0)\}$; 
$01=\{i: (Y_{i1}(0)=0, Y_{i1}(1)=1)\}$;
$10=\{i: (Y_{i1}(0)=1, Y_{i1}(1)=0)\}$; 
and $11=\{i: (Y_{i1}(0)=1, Y_{i1}(1)=1)\}$. 
Let $G_i$ denote the principal stratum membership for unit $i$, with $i=(1, \ldots, n$), then $G_i\equiv \left(Y_{i1}(0), Y_{i1}(1)\right) \in \{00,01,10,11\}$.
It is worth noting that  a binary version of a multi-valued or continuous lagged outcome is still a measure of the same substantive quantity as the final outcome  but measured at a previous time-point, and allows us to captures all the conceptual issues without involving complex model structures. In our illustrative example, the intermediate outcome is an indicator variable taking on value one if a firm hires new staff between the assignment to the first treatment and the assignment to the second treatment. Therefore, for example, principal stratum $11$ includes firms that would hire new staff irrespective of their treatment assignment at time $t=1$ (see Section~\ref{sec:Case} for further details).

Principal stratum membership $G_i$ is not affected by treatment assignment at time $t=1$, $W_{i1}$, so it only reflects characteristics of unit $i$.  Unfortunately we cannot, in general, observe the principal stratum which a unit belongs to, because principal strata are defined by the joint values of observed and missing intermediate outcomes. Therefore, principal strata can be viewed as a representation of the latent unobserved structure that may influence the decision to participate in the treatment at a future time point.

Based on principal stratification, we introduce a Latent Sequential Ignorability (LSI) assumption, where the word \textit{latent} indicates that treatment assignment is conditionally independent on future potential outcomes conditionally on pre-treatment covariates, past treatments and the latent indicator for principal stratum membership. 
In other words, LSI is a form of latent ignorability \citep{FrangRubin:1999}, in that it conditions on variables that are (at least partially) unobserved or latent. Formally:
\begin{assumption}\label{ass:LSI_AM} Latent Sequential Ignorability (LSI) 
\begin{eqnarray*}
& Pr\left(\bW_{i} \mid \bX_i, Y_{i1}(0), Y_{i1}(1), Y_{i2}(0,0), Y_{i2}(1,0), Y_{i2}(0,1), Y_{i2}(1,1)\right)=&\\
& Pr\left(W_{i1} \mid \bX_i\right) \times 
Pr(W_{i2} \mid W_{i1}, Y_{i1}(0), Y_{i1}(1), \bX_i).&
\end{eqnarray*}
\end{assumption}
LSI is a relaxed version of SIL (Assumption \ref{ass:SI_AM}): SIL implies LSI, therefore SIL is a stronger assumption.
 We formally show the relationship between SIL and LSI in the Appendix.
The proof proceeds by first showing that SIL can be equivalently formulated as follows
	\begin{eqnarray}
	Pr\left(W_{i1}\mid \bX_i, Y_{i1}(0), Y_{i1}(1), Y_{i2}(0,0), Y_{i2}(1,0), Y_{i2}(0,1), Y_{i2}(1,1)\right)&=&Pr\left(W_{i1} \mid \bX_i\right) \label{eq:SIL1}\\
	Pr\left(W_{i2} \mid \bX_i,W_{i1}, Y_{i1}(0), Y_{i1}(1), Y_{i2}(0,0), Y_{i2}(1,0), Y_{i2}(0,1), Y_{i2}(1,1) \right) &=& Pr\left(W_{i2} \mid \bX_i, W_{i1}, Y_{i1}^{obs}\right)\label{eq:SIL2}
	\end{eqnarray} 
and  LSI can be equivalently formulated as follows
\begin{eqnarray}
Pr\left(W_{i1}\mid \bX_i, Y_{i1}(0), Y_{i1}(1), Y_{i2}(0,0), Y_{i2}(1,0), Y_{i2}(0,1), Y_{i2}(1,1)\right)&=&Pr\left(W_{i1} \mid \bX_i\right) \label{eq:LSI1}\\
Pr\left(W_{i2} \mid \bX_i,W_{i1}, Y_{i1}(0), Y_{i1}(1), Y_{i2}(0,0), Y_{i2}(1,0), Y_{i2}(0,1), Y_{i2}(1,1) \right) &=& Pr\left(W_{i2} \mid \bX_i, W_{i1}, Y_{i1}(0), Y_{i1}(1)\right)\label{eq:LSI2}
\end{eqnarray} 
 As we can easily see, Equations \eqref{eq:SIL1} and \eqref{eq:SIL2} imply Equations \eqref{eq:LSI1} and \eqref{eq:LSI2}, and thus we have that SIL implies LSI.

The formulation of LSI  through Equations \eqref{eq:LSI1} and \eqref{eq:LSI2} makes it also clear the critical difference between SI (Assumption \ref{ass:SI}) and LSI. Although SI and LSI both assume that the assignment mechanism at time $t=1$ is ignorable given the set of observable variables, $\bX_i$ (see Equation \eqref{eq:SI1} and Equation \eqref{eq:LSI1}), SI and LSI impose different restrictions on  the assignment mechanism at time $t=2$: standard sequential ignorability implies that the assignment mechanism is ignorable given the observable past history, whereas LSI requires that it is ignorable given the observable past history  and the missing intermediate outcomes.
 
LSI implies that
\begin{eqnarray*}
E\left[Y_{i2}(w_1, w_2) \mid \bX_i \right] = 
\int
E\left[Y^{obs}_{i2} \mid W_{i1}=w_1, W_{i2}=w_2, G_i=g, \bX_i\right] d\, F_{G_i\mid  \bX_i}(g),
\end{eqnarray*}
where $F_{G_{i}\mid  \bX_i}(\cdot)$ is the conditional cumulative distribution function of the principal stratum membership, $G$, given  pre-treatment covariates. 
Therefore if principal stratum membership  were observed, under LSI within cells defined by the covariates, 	$E\left[Y_{i2}(w_1, w_2) \mid \bX_i \right]$ could be derived as the weighted average of the means of the observed outcome for units with $W_{i1}=w_1$ and $W_{i2}=w_2$  across principal strata with weights that depends on the conditional distribution of principal strata given covariates. In practice, principal stratum membership is generally unobserved, therefore inference under LSI raises non trivial challenges (see Section \ref{sec:BI} for details on inference under LSI).

\section{Assessing Sequential Ignorability through Latent Sequential Ignorability}
\label{sec:PrW2}
In this section we investigate the role of LSI (Assumption \ref{ass:LSI_AM}) in causal inference for sequential treatment.
Let first consider the relationship between SIL (Assumption \ref{ass:SI_AM}) and LSI (Assumption \ref{ass:LSI_AM}). LSI is a relaxed version of SIL and for this reason SIL can be viewed as a special case of LSI. Therefore, in order to compare SIL with LSI and to investigate which one is more appropriate for a given problem at hand, we rely on the relationship between SIL and LSI when SIL holds.

Under SIL, treatment assignment at $t=2$ does not depend on the missing intermediate potential outcomes, implying that treatment assignment probabilities are homogeneous across some principal strata, conditionally on the treatment assigned at $t=1$ and covariates.
Specifically, SIL implies that the assignment probabilities of $W_{i2}$ in principal strata sharing  the same value for the observed intermediate  outcome that is, the intermediate outcome  under the treatment assigned at time 1, are the same. 
Formally, under SIL, for  each $w_1=0,1$ and $y_1=0,1$, we have
\[
Pr(W_{i2}| \bX_i, W_{i1}=w_1, Y_{i1}^{obs}=y_1, Y_{i1}^{mis})  = Pr(W_{i2}| \bX_i, W_{i1}=w_1, Y_{i1}^{obs}=y_1).
\]
Therefore, if SIL holds we have:
\begin{equation}
\label{eq:equal}
\begin{array}{lcl}
Pr(W_{i2}| \bX_i, W_{i1}=0, G_{i}=00) &=& Pr(W_{i2}| \bX_i, W_{i1}=0, G_{i}=01), \\
Pr(W_{i2}| \bX_i, W_{i1}=0, G_{i}=10) &=& Pr(W_{i2}| \bX_i, W_{i1}=0, G_{i}=11), \\
Pr(W_{i2}| \bX_i, W_{i1}=1, G_{i}=00) &=& Pr(W_{i2}| \bX_i, W_{i1}=1, G_{i}=10), \\
Pr(W_{i2}| \bX_i, W_{i1}=1, G_{i}=01) &=& Pr(W_{i2}| \bX_i, W_{i1}=1, G_{i}=11). 
\end{array}
\end{equation}

Under SI (Assumption \ref{ass:SI}) the assignment probabilities  of $W_{i2}$ only depend on the observed intermediate outcomes conditionally on the treatment assigned at $t=1$ and covariates:  $Pr(W_{i2}| \bX_i, W_{i1}=w_1, Y_{i1}^{obs})$, therefore they can be ignored in drawing inference on the causal effects of interest. 
If SI does not hold, but LSI holds, ignoring the assignment probabilities  of $W_{i2}$, $Pr(W_{i2}| \bX_i, W_{i1}=w_1, G_{i}=g)$, does not, in general, lead to a valid analysis. 
This result suggests that we can investigate the robustness of the estimated causal effects  with respect to   violations of the sequential ignorability assumptions, using the assignment probabilities under LSI, $Pr(W_{i2}| \bX_i, W_{i1}=w_1, G_{i}=g)$, as sensitivity parameters. 

If principal strata encode  characteristics of the units that are  associated with the treatment assigned at time $t=2$ and possibly with the final outcome, i.e., LSI holds but  neither SIL nor SI holds, inference under LSI is expected to show evidence against at least one of the equalities in Equation~\eqref{eq:equal}, and SI/SIL and LSI are expected to lead to substantially different inferential conclusions on the causal effects of interest.  Conversely, if  we find that treatment assignment probabilities are homogeneous across principal strata according to the equalities in Equation~\eqref{eq:equal}, then causal  inference under sequential ignorability is more defensible. 

In this sense, LSI naturally provides a framework for  sensitivity analysis with respect to violations of  sequential ignorability: looking at the inferential results on the assignment probabilities under LSI we can get some insight on the plausibility of the sequential ignorability assumptions.
 Specifically we use a Bayesian approach to inference by proposing to look at the posterior distributions of  the treatment  assignment probabilities at time $t=2$ under LSI. If we find substantial overlap between the posterior  distributions of  the treatment  assignment probabilities at time $t=2$ across specific pairs of strata according to the equalities in Equation~\eqref{eq:equal}, 
 then we consider reasonable to conduct causal inference using  methods appropriate under sequantial ignorability assumptions. In this scenario LSI can be merely used as    a tool for  sensitivity analysis.
 Conversely if  Bayesian inference under LSI shows   evidence against some equality in Equation~\eqref{eq:equal}, leading to posterior distributions of the corresponding treatment  assignment probabilities at time $t=2$ rather apart, then   sequantial ignorability assumptions are deemed implausible and we argue that causal inference under LSI is more defensible.
 
For instance, in our illustrative study, Equation~\eqref{eq:equal} implies that we can assess the plausibility of sequential ignorability assumptions  by comparing  the posterior distributions of the probabilities of receiving a loan in the second treatment period (i.e., 2005 to 2007) obtained under LSI across pairs of principal strata. For example, according to the first equality in Equation~\eqref{eq:equal},  we compare the posterior distributions of the conditional probabilities of receiving a loan in the second treatment period between the following two groups of firms with the same background characteristics that did not receive any loan in the first treatment period: (a) firms in principal stratum $00$, which would not hire new staff irrespective of the treatment received at time $t = 1$ (between 2002 and 2004); and (b) firms in principal stratum $01$, which would hire new staff if granted at time $t = 1$ but would not hire if not granted at time $t = 1$. If we find some evidence against some equality in Equation~\eqref{eq:equal}, we can argue that firms' decision to apply for a loan in the second treatment period depends on unobserved confounders related to the principal strata defined by the firms' hiring policy above and beyond firms' observed history. Thus,  inference on the average causal effects on the number of employees is deemed as more reliable under LSI than under sequential ignorability assumptions.
 This framework for sensitivity analysis is in line with the existing approaches in the literature to sensitivity analysis with respect to violations of the unconfoundness assumption, usually made in single time observational studies  \cite[][]{RosenbaumRubin:1983, Rosenbaum:1987, Imbens:2003, NIM:2008, Ding:VanderWeele:2016}. There the robustness of the estimated causal effects with respect to the unconfoundness assumption is generally assessed focusing on its violations due to the presence of unobserved covariates that are correlated both with the potential outcomes and with the treatment indicator. In those settings, sensitivity parameters are  quantities  characterizing the distribution of the hypothetical unobserved covariates and their association with the  potential outcomes and with the treatment indicator, but they do not generally have a substantial meaning.
There is  no  evidence in the data about the association between the hypothetical unobserved covariates and the potential outcomes and the treatment indicator; hypothetical unobserved confounders  are usually defined using subject matter knowledge, which may be debatable. Conversely, longitudinal data coupled with the principal stratification framework 
	provide valuable information on the presence of possible unmeasured confounders breaking sequential ignorability assumptions. 
 In this framework, sensitivity parameters are meaningful quantities with a direct interpretation: they are  assignment probabilities for specific sub-population of units.

 Principal stratification can be viewed as a coarsened representation of (post-treatment) unmeasured confouders, which may be  binary, categorical or continuous.
%Specifically, the (basic) principal stratification with respect to a binary post-treatment intermediate outcome  classifies units into 4 latent groups, that is, the indicator for principal stratum membership, $G_i$,  is a categorical variable with four levels.
It is reasonable to believe that if there existed unmeasured  confounders that are related to $W_{i2}$ and $Y_{i2}$, principal stratum membership, $G_i$, which is a categorical variable with four levels,  would depend on the distribution of those unmeasured  confounders. Therefore in some sense we can view the principal stratification  approach to sensitivity analysis 
with respect to violations of  sequential ignorability assumptions as non-parametric.

\section{Inference} 
\label{sec:Inference}
Under SI and SIL average causal effects are point identified, i.e., they can be expressed as known function of the distribution of the observed data, since different effect values cannot correspond to the same distribution of the observables.
Therefore, ideally, we could estimate average treatment effects non-parametrically. In practice, data are often sparse and high dimensional, and model assumptions are usually introduced. 
Methods usually applied to estimate causal effects of longitudinal treatments under SI (Assumption \ref{ass:SI}) include the G-computation algorithm formula \citep{Robins:1986}, inverse probability of treatment weighting estimation  of marginal structural  models \citep{Robins:1989}, and G-estimation of structural nested models \citep{Robins:1999}. 
The three methods would give identical estimates of the treatment effects if  a non-parametric approach to inference or saturated marginal structural models/structured nested models were used, but under model assumptions they generally provide different estimates, depending on the specific parametric assumptions that are introduced. 
The G-formula requires to specify many models, often raising model-compatibility issues. Marginal structural models (MSMs) and structured nested models, which have received increasing attention in the last years, require to specify models for marginal potential outcomes ($Y_2(w_1, w_2)$ for each $(w_1,w_2) \in \{0,1\}^2$ in our setting) and for the causal effects, which may assume, e.g., constant treatment effects, additivity and so on. Moreover  inferential methods based on inverse probability of treatment weighting require to also specify a model for the probability of treatment.  
These assumptions may be critical because model misspecification  may lead to biased estimates of the treatment effects even if  the identifiability conditions hold. 
%\cite{RobinsHernan:2008}

Under LSI the average causal effects are generally not non-parametrically point identified, due to the latent nature of the  principal strata. In our setting, we can only observe four groups based on  the treatment actually received at time $t=1$, $W_{i1}$, and the observed value of the intermediate outcome,  $Y_{i1}^{obs}$, and each of them comprises a mixture of two principal strata, as shown in the last two columns in  Table~\ref{tab:ObsGr}.
  Therefore under LSI causal effects in Equation \eqref{eq:ATE} are only partially identified. 
\begin{table}
\caption{\label{tab:ObsGr} Group classification based on observed data $O(W_{i1}, Y_{i1}^{obs})$, associated data pattern and latent principal strata.}
\centering
\begin{tabular}{l c c | c c}
\hline
Observed group &  & & \multicolumn{2}{c}{Latent group} \\
$O(W_{i1}, Y_{i1}^{obs})$ & $W_{i1}$ & $Y_{i1}^{obs}$ & \multicolumn{2}{c}{$G_{i}$} \\
\hline
$O(0,0)$ & 0 & 0 & 00 & 01 \\
$O(0,1)$ & 0 & 1 & 10 & 11 \\
$O(1,0)$ & 1 & 0 & 00 & 10 \\
$O(1,1)$ & 1 & 1 & 01 & 11 \\
\hline
\end{tabular}
\end{table}
In the principal stratification literature,  structural or modeling assumptions are typically invoked  to deal with identification issues \cite[e.g.,][]{ImbensRubin:1997, MatteiMealli:2007, SchwartzLiMealli:2011}. Monotonicity and exclusion restriction assumptions, usually used in experimental studies with noncompliance, may be questionable in longitudinal settings.
Depending on the substantive empirical setting, other structural or modeling assumptions can be introduced. In this paper we prefer to avoid structural assumptions, which may make the comparison between SI/SIL and LSI unfair or strongly depending on some specific assumption, and we opt for a  model-based approach for inference.
 
Following the literature on principal stratification, models for potential outcomes are specified conditional on covariates and principal strata (see Section~\ref{sec:BI} for further details). 
Again, distributional assumptions may be critical. Nevertheless in our opinion this model-based approach is very flexible, and in some settings model assumptions on the conditional distributions of potential outcomes may be less demanding than model assumptions on the marginal distributions of potential outcomes and on the causal effects. 
In order to make the comparison between SI/SIL and LSI as fair as possible, the same model-based approach is used under SI/SIL, although we will also employ G-methods under SI. An advantage of the model-based approach is that it allows us to directly get information on the heterogeneity of the effects with respect to principal strata both under SI/SIL and LSI.

\subsection{Bayesian Inference}
\label{sec:BI}
We adopt a Bayesian approach to inference, which is particularly suitable for model-based causal inference   and appears to be a natural and appealing inferential approach to make comparisons between SI/SIL and LSI on the same ground.  The Bayesian perspective   is particularly appropriate for addressing problems of causal inference because it treats the uncertainty in the missing potential outcomes in the same way that it treats the uncertainty in the unknown parameters. A Bayesian approach explicitly deals with the different sources of uncertainty, treating them separately. Also in a Bayesian framework, we can be formally clear about the role played by the treatment assignment mechanism and the complications that
raise in drawing inference for sequential treatments under LSI \cite[][]{Rubin:1978, ImbensRubin:1997}.
From a Bayesian perspective, all inferences are based on the posterior distribution of the causal estimands, defined as functions of observed and unobserved potential outcomes, or sometimes as functions of model parameters \citep{Rubin:1978}.  Because with proper prior distributions, posterior distributions are always proper, from a Bayesian perspective, there is no conceptual difference between fully and partially   identified parameters \cite[][]{Gustafson:2010}. 
In Bayesian inference, the key difference between fully and partially identified parameters concerns their limiting distribution: 
as the sample size goes to
infinity the support of the marginal posterior distribution of a fully identified parameter converges to a single value, whereas the support of the marginal posterior distribution of a partially identified parameter
converges to the identification region, which is a set with cardinality smaller than the cardinality of the corresponding prior support, but larger than one \cite[][]{Gustafson:2010}. The shape
of the limiting distribution of a  partially identified parameter may provide valuable information on the parameters of interest and the choice of the prior may affect the informativeness of the shape of the limiting distribution \cite[][]{Gustafson:2010, Gustafson:2014}. Distributional assumptions may help identification, making inference less sensitive to the choice of the specification of the prior distributions. 
Here we assume that under LSI potential outcomes
for the final response variable, $Y_{i2}(w 1 , w 2)$, are normally distributed conditional on principal
stratum membership, therefore we end up with dealing with finite mixture of Normal distributions, which are identifiable \cite[e.g.][]{McLachlanPeel:2000}. 
Nevertheless we also need to account that we have information on a
finite sample of units and, with finite samples, posterior distributions of partially identified
parameters usually have substantial region of flatness. This feature, which may be shared
also with some fully identified parameters, is called weakly identifiability \cite[e.g.,][]{ImbensRubin:1997}. In Bayesian analysis of weakly identifiable models, investigating the robustness
of the results with respect to the specifications of the priors might be worthwhile to make
inferences more reliable. Therefore we also  investigate the robustness of the simulation results with respect to the specification of prior distributions.

Bayesian inference considers the observed values to be realizations of random variables and the missing values to be unobserved random variables, starting from  the joint probability distributions of all random variables for all units: 
\[
p(\bY_1(0), \bY_1(1), \bY_2(0,0), \bY_2(0,1), \bY_2(1,0), \bY_2(1,1), \bW_1, \bW_2, \bX).
\]
We assume this distribution is unit exchangeable, that is, invariant under a permutation of the indexes, then de Finetti's theorem \citep{deFinetti:1937, deFinetti:1964}  implies that there exists a vector of parameters $\btheta$, which is a random variable itself, with prior distribution $p(\btheta)$, such that $\bY_1(0), \bY_1(1), \bY_2(0,0), \bY_2(0,1), \bY_2(1,0), \bY_2(1,1), \bW_1, \bW_2$ and $\bX$ consist of independent and identically distributed random variables given $\btheta$. Thus,  
\begin{eqnarray}
\label{eq:deFin_X}
\lefteqn{p(\bY_2(0,0), \bY_2(0,1), \bY_2(1,0), \bY_2(1,1), \bY_1(0), \bY_1(1), \bW_1, \bW_2,  \bX) = } \nonumber \\
&  \int \prod_i p(Y_{i2}(0,0), Y_{i2}(0,1), Y_{i2}(1,0), Y_{i2}(1,1), Y_{i1}(0), Y_{i1}(1), W_{i1}, W_{i2},  \bX_i \mid \btheta) p(\btheta) d\btheta, \nonumber
\end{eqnarray}
and the posterior distribution of $\btheta$ can be written as
\begin{eqnarray}
\label{eq:postTheta}
\lefteqn{p(\btheta \mid \bY_2^{obs}, \bY_1^{obs}, \bW_1, \bW_2, \bX)\propto  }  \nonumber \\
&&\!\!\!\!\!\! \!\!\!  p(\btheta) \times \int  \! \! \! \int  \prod_i  p(Y_{i2}(0,0), Y_{i2}(0,1), Y_{i2}(1,0), Y_{i2}(1,1), Y_{i1}(0), Y_{i1}(1), W_{i1}, W_{i2},  \bX_i \mid \btheta)  d\bY_{i2}^{mis} dY_{i1}^{mis}= \nonumber \\
&& \!\!\!\!\!\! \!\!\!p(\btheta)  \times  \nonumber \\&&\!\!\!\!\!\! \!\!\!
\int \! \! \! \int  \prod_i  \Bigl[ p(\bX_i \mid \btheta) \! \times\!
p( Y_{i1}(0), Y_{i1}(1) | \bX_i; \btheta) \times  p(Y_{i2}(0,0), Y_{i2}(0,1), Y_{i2}(1,0), Y_{i2}(1,1) \mid  Y_{i1}(0), Y_{i1}(1), \bX_i; \btheta) \times \nonumber \\&&
 \Bigl.
 \qquad p(W_{i1} \mid Y_{i2}(0,0), Y_{i2}(0,1), Y_{i2}(1,0), Y_{i2}(1,1), Y_{i1}(0), Y_{i1}(1),   \bX_i; \btheta) \times \nonumber\\&&
\qquad p( W_{i2} \mid Y_{i2}(0,0), Y_{i2}(0,1), Y_{i2}(1,0), Y_{i2}(1,1), Y_{i1}(0), Y_{i1}(1), \bX_i, W_{i1}; \btheta) 
d\bY_{i2}^{mis} dY_{i1}^{mis}\Bigr]. \nonumber
\end{eqnarray}

The assumptions on the assignment mechanism are crucial to draw inference on the causal estimands. 
Under latent sequential ignorability (Assumption \ref{ass:LSI_AM}),
within cells defined by the values of pre-treatment
variables $\bX_i$, the treatment at time $t=1$  is assigned independently of the relevant post-treatment variables, $Y_{i1}(w_1)$ and $Y_{i2}(w_1,w_2)$, $w_1=0,1$, $w_2=0,1$, and 
the treatment at time $t=2$  is assigned independently of the final potential outcomes, $Y_{i2}(w_1,w_2)$, $w_1=0,1$, $w_2=0,1$,  conditional on the treatment assigned at time $t=1$, $W_{i1}$, and the principal strata defined by  $(Y_{i1}(0),Y_{i1}(1))$. Therefore, under LSI the posterior distribution of $\btheta$ becomes
\begin{eqnarray}
\label{eq:postTheta_LSI}
\lefteqn{p(\btheta \mid \bY_2^{obs}, \bY_1^{obs}, \bW_1, \bW_2, \bX) \propto
 p(\btheta)  \times    }  \nonumber \\ 
 & &\!\!\!\!\!\!\!\! \int \! \! \int  \prod_i  \bigl[ p(\bX_i \mid \btheta) \!\times\! p( Y_{i1}(0), Y_{i1}(1) \mid \bX_i; \btheta) \!\times\!  p(Y_{i2}(0,0), Y_{i2}(0,1), Y_{i2}(1,0), Y_{i2}(1,1) \mid  Y_{i1}(0), Y_{i1}(1), \bX_i; \btheta) \!\times\!
 \nonumber \\
 &&  p(W_{i1} \mid  \bX_i; \btheta) \!\times\! p( W_{i2} \mid Y_{i1}(0), Y_{i1}(1), \bX_i, W_{i1}; \btheta) 
d\bY_{i2}^{mis} dY_{i1}^{mis}\bigr]. 
\end{eqnarray}
Equation~\eqref{eq:postTheta_LSI} further simplifies under SIL (Assumption \ref{ass:SI_AM}), which implies that the treatment 
at time $t=2$  is assigned independently of both missing intermediate potential outcomes  $Y^{mis}_{i1}$ and  final potential outcomes, $Y_{i2}(w_1,w_2)$, $w_1=0,1$, $w_2=0,1$,  conditional on  the pre-treatment variables, $\bX_i$, the treatment assigned at time $t=1$, $W_{i1}$, and the past observed potential outcomes, $Y^{obs}_{i1}$:
\begin{eqnarray}
\label{eq:postTheta_SI}
\lefteqn{p(\btheta \mid \bY_2^{obs}, \bY_1^{obs}, \bW_1, \bW_2, \bX) \propto
 p(\btheta)  \times    } \nonumber \\ 
 & &\!\!\!\!\!\!\!\! \int \! \! \int  \prod_i  \bigl[ p(\bX_i \mid \btheta) \!\times\! p( Y_{i1}(0), Y_{i1}(1) \mid \bX_i; \btheta) \!\times\!  p(Y_{i2}(0,0), Y_{i2}(0,1), Y_{i2}(1,0), Y_{i2}(1,1) \mid  Y_{i1}(0), Y_{i1}(1), \bX_i; \btheta) \!\times\!
 \nonumber \\&&  p(W_{i1} \mid   \bX_i; \btheta) \!\times\! p( W_{i2} \mid Y^{obs}_{i1}, \bX_i, W_{i1}; \btheta) 
d\bY_{i2}^{mis} dY_{i1}^{mis}\bigr]. 
\end{eqnarray}
The right hand of Equation \eqref{eq:postTheta_SI} is also the posterior distribution of $\btheta$  under SI (Assumption \ref{ass:SI}).
It is worth noting that, under the assumption that the parameters governing the distributions under the integral sign in Equations~\eqref{eq:postTheta_LSI} and \eqref{eq:postTheta_SI} are a priori distinct and independent from each other  \citep{Rubin:1978}, we can ignore  the distributions $p(\bX_i \mid \btheta_{X})$ and  $p(W_{i1} \mid   \bX_i; \btheta)$ in drawing Bayesian inference on the relevant estimands. If SI/SIL holds, Bayesian causal inference does not even require to model the distribution of the treatment at time $t=2$, $p( W_{i2} \mid Y^{obs}_{i1}, \bX_i, W_{i1}; \btheta)$ \cite[][]{Rubin:1978, Zajonc:2012}, although we decided to model it in the analyses below to better describe and discuss the  role of LSI and SI/SIL in longitudinal studies.

Throughout the article we assume that conditional on $\bX_i$, $G_i$ and $\btheta$, the four outcomes $Y_{i2}(0,0)$, $Y_{i2}(0,1)$, $Y_{i2}(1,0)$, $Y_{i2}(1,1)$ are independent. Data are not informative about the partial association structure between final potential outcomes, because  $Y_{i2}(0,0)$, $Y_{i2}(0,1)$, $Y_{i2}(1,0)$, $Y_{i2}(1,1)$ are never jointly observed, but the independence assumption has little inferential effect if we regard the $n$ units in the study as a random sample from a super-population and we focus on super-population causal estimands that do not depend on the  association structure between the final potential outcomes.
Indeed, the causal estimands of primary interest here, the average causal effects in Equation~\eqref{eq:ATE} are super-population causal estimands, which are free of the  association structure between the final potential outcomes \cite[][Chapter 6, pp. 98-101]{ImbensRubin:1997, ImbensRubin:2015}.

Let $O(W_{i1}, Y_{i1}^{obs}, W_{i2})$ denote the observed group defined by the observed variables $W_{i1}$, $Y_{i1}^{obs}$, and  $W_{i2}$, and recall that $G_i \equiv (Y_{i1}(0), Y_{i1}(1)) \in \{00,01,10,11\}$. 
 Define
$\pi_{ig} = p( G_i=g \mid \bX_i; \btheta)$
and $h_{ig}^{w_1}= p(W_{i2} = 1 \mid  G_i=g, \bX_i, W_{i1}=w_1; \btheta)$ for $g=00,01,10,11$, $w_1=0,1$;
and let $f_{ig}^{w_1,w_2}(y_2)= f_{Y_{i2}(w_1,w_2) \mid  G_i=g, \bX_i; \btheta}(y_2)$ be the probability mass/density function  of $Y_{i2}(w_1,w_2) \mid  G_i=g, \bX_i; \btheta$,  $g=00,01,10,11$, $w_1=0,1$, $w_2=0,1$. Then performing the integration in Equation~\eqref{eq:postTheta_LSI}, under LSI the posterior distribution of $\btheta$ given the observed data
can be written as follows:
{\small
	\begin{eqnarray}\label{eq:postTheta_LSIb}
\lefteqn{p(\btheta | \bY_2^{obs}, \bY_1^{obs}, \bW_1, \bW_2, \bX) \propto
 p(\btheta)  \times }  \nonumber \\ & &
\hspace{-0.75cm} \prod_{i \in O(0,0,0) }
  \left[\pi_{i00}  (1-h_{i00}^{0}) f_{i00}^{0,0}(Y_{i2}^{obs})+
\pi_{i01}  (1-h_{i01}^{0}) f_{i01}^{0,0}(Y_{i2}^{obs})\right] \times
 \prod_{i \in O(0,0,1)}
  \left[\pi_{i00}  h_{i00}^{0} f_{i00}^{0,1}(Y_{i2}^{obs})+
\pi_{i01}  h_{i01}^{0} f_{i01}^{0,1}(Y_{i2}^{obs})\right] \times \nonumber\\ & &
\hspace{-0.75cm} 
 \prod_{i \in O(0,1,0) }
  \left[\pi_{i10}  (1-h_{i10}^{0}) f_{i10}^{0,0}(Y_{i2}^{obs})+
\pi_{i11}  (1-h_{i11}^{0}) f_{i11}^{0,0}(Y_{i2}^{obs})\right] \times
 \prod_{i \in O(0,1,1)}
  \left[\pi_{i10}  h_{i10}^{0} f_{i10}^{0,1}(Y_{i2}^{obs})+
\pi_{i11}  h_{i11}^{0} f_{i11}^{0,1}(Y_{i2}^{obs})\right] \times \nonumber\\ & &
\hspace{-0.75cm} 
% %
 \prod_{i \in O(1,0,0) }
  \left[\pi_{i00}  (1-h_{i00}^{1}) f_{i00}^{1,0}(Y_{i2}^{obs})+
\pi_{i10}  (1-h_{i10}^{1}) f_{i10}^{1,0}(Y_{i2}^{obs})\right] \times
 \prod_{i \in O(1,0,1)}
  \left[\pi_{i00}  h_{i00}^{1} f_{i00}^{1,1}(Y_{i2}^{obs})+
\pi_{i10}  h_{i10}^{1} f_{i10}^{1,1}(Y_{i2}^{obs})\right] \times \nonumber\\ & &
\hspace{-0.75cm}
 \prod_{i \in O(1,1,0) }
  \left[\pi_{i01}  (1-h_{i01}^{1}) f_{i01}^{1,0}(Y_{i2}^{obs})+
\pi_{i11}  (1-h_{i11}^{1}) f_{i11}^{1,0}(Y_{i2}^{obs})\right] \times
 \prod_{i \in O(1,1,1)}
  \left[\pi_{i01}  h_{i01}^{1} f_{i01}^{1,1}(Y_{i2}^{obs})+
\pi_{i11}  h_{i11}^{1} f_{i11}^{1,1}(Y_{i2}^{obs})\right]. % \nonumber\\ & &
\end{eqnarray}
}
Therefore model-based Bayesian inference under LSI requires to specify three models:  $(1)$ the model for principal strata conditional on covariates, $\pi_{ig}$;$(2)$ the model for treatment assigned at time $t=2$ conditional on principal strata, past treatment and covariates, $h_{ig}^{w_1}$; and $(3)$ the model for final potential outcomes conditional on principal strata and covariates, $f_{ig}^{w_1,w_2}$.

Let $h_{iy_1}^{w_1}= p(W_{i2} = 1 \mid Y_{i1}(w_1)=y_1, \bX_i, W_{i1}=w_1; \btheta)$, $y_1=0,1$, $w_1=0,1$. 
Performing the integration in Equation~\eqref{eq:postTheta_SI}, under SI/SIL the posterior distribution of $\btheta$ given the observed data can be written as follows:
\begin{eqnarray}\label{eq:postTheta_SIb}
\lefteqn{p(\btheta | \bY_2^{obs}, \bY_1^{obs}, \bW_1, \bW_2, \bX) \propto
 p(\btheta)  \times } \nonumber \\ & &
 \prod_{i \in O(0,0,0) }
 (1-h_{i0}^{0}) \left[\pi_{i00}   f_{i00}^{0,0}(Y_{i2}^{obs})+
\pi_{i01}   f_{i01}^{0,0}(Y_{i2}^{obs})\right] \times
 \prod_{i \in O(0,0,1)}
 h_{i0}^{0} \left[\pi_{i00}   f_{i00}^{0,1}(Y_{i2}^{obs})+
\pi_{i01}   f_{i01}^{0,1}(Y_{i2}^{obs})\right] \times \nonumber\\ & &
 \prod_{i \in O(0,1,0) }
(1-h_{i1}^{0})   \left[\pi_{i10}  f_{i10}^{0,0}(Y_{i2}^{obs})+
\pi_{i11}  f_{i11}^{0,0}(Y_{i2}^{obs})\right] \times
 \prod_{i \in O(0,1,1)}
h_{i1}^{0}  \left[\pi_{i10}   f_{i10}^{0,1}(Y_{i2}^{obs})+
\pi_{i11}  f_{i11}^{0,1}(Y_{i2}^{obs})\right] \times \nonumber\\ & &
% %
 \prod_{i \in O(1,0,0) }
(1-h_{i0}^{1})  \left[\pi_{i00}   f_{i00}^{1,0}(Y_{i2}^{obs})+
\pi_{i10}   f_{i10}^{1,0}(Y_{i2}^{obs})\right] \times
 \prod_{i \in O(1,0,1)}
h_{i0}^{1}  \left[\pi_{i00}   f_{i00}^{1,1}(Y_{i2}^{obs})+
\pi_{i10}   f_{i10}^{1,1}(Y_{i2}^{obs})\right] \times \nonumber\\ & &
 \prod_{i \in O(1,1,0) }
(1-h_{i1}^{1})  \left[\pi_{i01}   f_{i01}^{1,0}(Y_{i2}^{obs})+
\pi_{i11}   f_{i11}^{1,0}(Y_{i2}^{obs})\right] \times
 \prod_{i \in O(1,1,1)}
h_{i1}^{1}  \left[\pi_{i01}   f_{i01}^{1,1}(Y_{i2}^{obs})+
\pi_{i11}   f_{i11}^{1,1}(Y_{i2}^{obs})	\right].  \nonumber
\\ & &
\end{eqnarray}
  Now, define  $\pi_{iw_1}=p(Y_{i1}(w_1)= 1 \mid \bX_i, \btheta)$, $w_1=0,1$ and $f_{iy_1}^{w_1,w_2}(y_2)= f_{Y_{i2}(w_1,w_2) = y_2 \mid Y_{i1}(w_1)=y_1, \bX_i; \btheta}(y_2)$, $y_1=0,1$, $w_1=0,1$, $w_2=0,1$. Then, taking the sums in the brackets on the right hand of Equation \eqref{eq:postTheta_SIb}, that is, marginalizing over the missing intermediate outcome, we have that 
	$\pi_{i0y_1}   f_{i0y_1}^{1,w_2}(y_2)+
	\pi_{i1y_1}   f_{i1y_1}^{1,w_2}(y_2) = \pi_{i1}^{y_1} (1-\pi_{i1})^{1-y_1}f_{iy_1}^{1,w_2}(y_2)$
	and 	$\pi_{iy_10}   f_{iy_10}^{0,w_2}(y_2)+
	\pi_{iy_11}   f_{iy_11}^{0,w_2}(y_2) = \pi_{i0}^{y_1} (1-\pi_{i0})^{1-y_1}f_{iy_1}^{0,w_2}(y_2)$
	. Therefore under SI/SIL, the posterior distribution of $\btheta$ given the observed data in Equation~\eqref{eq:postTheta_SIb} can be also written as follows:
	\begin{eqnarray}\label{eq:postTheta_SIc}
	\lefteqn{p(\btheta | \bY_2^{obs}, \bY_1^{obs}, \bW_1, \bW_2, \bX) \propto
		p(\btheta)  \times } \nonumber \\ & &
	\prod_{i \in O(0,0,0) }
	(1-h_{i0}^{0}) (1-\pi_{i0}) f_{i0}^{0,0}(Y_{i2}^{obs}) \times
	\prod_{i \in O(0,0,1)}
	h_{i0}^{0} (1-\pi_{i0})   f_{i0}^{0,1}(Y_{i2}^{obs}) \times \nonumber\\ & &
	\prod_{i \in O(0,1,0) }
	(1-h_{i1}^{0})   \pi_{i0}  f_{i1}^{0,0}(Y_{i2}^{obs}) \times 
	\prod_{i \in O(0,1,1)}
	h_{i1}^{0}  \pi_{i0}   f_{i1}^{0,1}(Y_{i2}^{obs}) \times \nonumber\\ & &
	% % 
	\prod_{i \in O(1,0,0) }
	(1-h_{i0}^{1})  (1-\pi_{i1})   f_{i0}^{1,0}(Y_{i2}^{obs}) \times
	\prod_{i \in O(1,0,1)}
	h_{i0}^{1}  (1-\pi_{i1})   f_{i0}^{1,1}(Y_{i2}^{obs}) \times\nonumber\\ & &
	\prod_{i \in O(1,1,0) }
	(1-h_{i1}^{1})  \pi_{i1}   f_{i1}^{1,0}(Y_{i2}^{obs})\times
	\prod_{i \in O(1,1,1)}
	h_{i1}^{1}  \pi_{i1}   f_{i1}^{1,1}(Y_{i2}^{obs}).
	\end{eqnarray}
Thus, on the basis of Equations~\eqref{eq:postTheta_SIb} and ~\eqref{eq:postTheta_SIc}, we can conduct model-based Bayesian inference under SI/SIL using two alternative model specifications.

Specifically, on the one hand Equation~\eqref{eq:postTheta_SIb} suggests to  use a specification similar to that we employ under LSI, specifying $(1)$ the model for principal strata conditional on covariates, $\pi_{ig}$;
$(2)$ the model for treatment assigned at time $t=2$ conditional on intermediate observed outcomes, past treatment and covariates, $h_{iy_1}^{w_1}$; and $(3)$ the model for final potential outcomes conditional on principal strata and covariates, $f_{ig}^{w_1,w_2}$.
This specification, which we refer to as specification \textit{SI-1}, implies that the only difference between Bayesian inference under LSI and SI-1 concerns the model for treatment assigned at time $t=2$, which depends on principal strata when LSI holds, but does only depend on the observed values of the intermediate outcome under SI/SIL.

On the other hand, on the basis of Equation~\eqref{eq:postTheta_SIc} we can model only the distributions for the observed data,  specifying  $(1)$ the model for intermediate observed outcomes conditional on covariates, $\pi_{iw_1}$; $(2)$ the model for treatment assigned at time $t=2$ conditional on intermediate observed outcomes, past treatment and covariates, $h_{iy_1}^{w_1}$; and $(3)$ the model for final potential outcomes conditional on    intermediate observed outcomes and covariates, $f_{iy_1}^{w_1,w_2}$. We refer to the  model specification based on Equation~\eqref{eq:postTheta_SIc} as specification \textit{SI-2}. 

Specification SI-2 reflects more closely the standard approaches to causal inference with longitudinal treatments under SI. Specification SI-1 may be preferable if we are interested in the heterogeneity of the effects across principal strata.

\section{Simulations}
\label{sec:Simulations}

In this Section we investigate the role of LSI (Assumption \ref{ass:LSI_AM})  and sequential ignorability assumptions (Assumption \ref{ass:SI} and Assumption \ref{ass:SI_AM}) using simulations. 
In our simulated experiment we set up two alternative scenarios in which both the data generating process and the assumptions underlying inference can vary. 
In the first scenario we generate data under sequential ignorability using a data generating process where both SI and SIL hold, while in the second scenario LSI   holds, but neither SI nor SIL holds.. Then we conduct Bayesian inference on the relevant causal estimands for each scenario under both LSI and SI, and in this latter case we use both the SI-1 and the SI-2 specifications.

In order to clearly assess the implications of the two alternative assumptions, LSI and SI/SIL, and investigate the robustness of the estimands to violations of SI/SIL, 
we focus on a simple setting: 
we assume 	either that LSI and SI/SIL hold without conditioning on the covariates  (and thus we can ignore the information on the covariates) or that
LSI and SI/SIL hold conditional on the observed covariates, but we are already within cells defined by observed pre-treatment variables. Indeed although LSI and SI/SIL  may be unrealistic without conditioning on the covariates, we can still  interpret the results by imaging that the analyses are conducted on a sub-population of units that is homogeneous with respect to the observed pretreatment variables.
We also consider relatively large sample sizes of $5000$ units to avoid (or, at least, reduce) sampling variability issues.

\subsection{Data generating processes}
\label{subsec:mod}

The true simulation models for all of the simulations are based on Equation~\eqref{eq:postTheta_LSIb} under LSI and  Equation~\eqref{eq:postTheta_SIb} under SI/SIL, which require to specify parametric models for principal strata ($\pi_{ig}$), 
the treatment assignment probabilities at time $t=2$  (either $h_{ig}^{w_1}$ or $h_{iy_1}^{w_1}$)
and the final outcome ( $f_{ig}^{w_1,w_2}$).
  The treatment at time 1, $W_{i1}$, is randomly assigned with probability   $h_i=0.5$, for all $i$.

The model for principal strata membership contains three conditional probit models,  defined using indicator variables $G_i(11)$,
 $G_i(00)$ and  $G_i(10)$  for whether unit $i$ belongs to principal stratum $11$, $00$ or $10$ (we use principal stratum $01$ as reference group):
\begin{eqnarray}\label{eq:Y1}
G_i(11) = 1 & \hbox{if} &  G_i^\ast(11) \equiv \alpha^{11}    +\epsilon_{i,11} \leq 0, \nonumber\\
G_i(00) = 1 & \hbox{if} &  G_i^\ast(11)>0 \, \hbox{ and } \, G_i(00)^\ast \equiv \alpha^{00}  
  + \epsilon_{i,00} \leq 0,\\
G_i(10) = 1 & \hbox{if} &  G_i^\ast(11)>0, \,  G_i(00)^\ast>0 \,  \hbox{ and } \, G_i(10)^\ast \equiv \alpha^{10}   + \epsilon_{i,10} \leq 0, \nonumber
\end{eqnarray}
where  $\epsilon_{i,11} \sim N(0,1)$, $\epsilon_{i,00} \sim N(0,1)$, and $\epsilon_{i,10} \sim N(0,1)$ independently.
Therefore
\begin{eqnarray*}
&\pi_{i11} =  1- \Phi \left( \alpha^{11}   \right), \qquad
\pi_{i00} = \Phi \left( \alpha^{11} \right) \left[ 1- \Phi  \left( \alpha^{00}  \right) \right], &\\
&
\pi_{i10}= \Phi \left( \alpha^{11}   \right) \Phi  \left( \alpha^{00} +\alpha_U^{00} U_i \right) \left[ 1- \Phi  \left( \alpha^{10}  \right) \right]&
\end{eqnarray*}
and $\pi_{i01} = 1-\sum_{g \in \{11, 00, 10\}} \pi_{ig} = \Phi(\alpha^{11}) \Phi(\alpha^{00}) \Phi(\alpha^{10})$, where $\Phi(\cdot)$ is the cumulative distribution function of the standard Normal distribution.

For the model of the treatment indicator at time $t=2$, $W_{i2}$, we use a probit specification. Under LSI, we assume the following probit model for the treatment assignment at time $t=2$:
\begin{eqnarray}\label{eq:W2_LSI}
W_{i2} = 1 & \hbox{if} &  W_{i2}^\ast \equiv \gamma_{w_1} + \gamma_{w_1}^{Y_1(0)} Y_{i1}(0)+ \gamma_{w_1}^{Y_1(1)} Y_{i1}(1) + \gamma_{w_1}^{Y_1(0)Y_1(1)}Y_{i1}(0)Y_{i1}(1)    + \epsilon_{i,W_2} > 0, 
\end{eqnarray}
where $\epsilon_{i,W_2} \sim N(0,1)$, $w_1 \in \{0,1\}$,  $Y_{i1}(0) \in  \{0, 1 \}$ and $Y_{i1}(1) \in  \{0, 1 \}$.
Under model \eqref{eq:W2_LSI}, the treatment assignment probabilities  at time $t=2$ are 
$$
h_{ig}^{w_1} = 
\begin{cases}
\Phi \left( \gamma_{w_1}\right)  & \hbox{ if }  W_{i1}=w_1  \hbox { and }  G_i=00;\\
\Phi \left(\gamma_{w_1} + \gamma_{w_1}^{Y_1(0)}\right)  & \hbox{ if } W_{i1}=w_1 \hbox { and }  G_i=10;\\
\Phi \left( \gamma_{w_1} +\gamma_{w_1}^{Y_1(1)}\right)  & \hbox{ if }  W_{i1}=w_1  \hbox { and } G_i=01;\\
\Phi \left( \gamma_{w_1} + \gamma_{w_1}^{Y_1(0)} + \gamma_{w_1}^{Y_1(1)}  + \gamma_{w_1}^{Y_1(0)Y_1(1)} \right)   & \hbox{ if }  W_{i1}=w_1  \hbox { and } G_i=11;\\
\end{cases}
$$
with $w_1=0,1$. 

Under SI/SIL, treatment assignment probabilities at time $t=2$ are free of the missing values for the intermediate outcome, either because $W_{i2}$ does not depends on the missing past potential outcomes or because only observed past potential outcomes enter the assignment mechanism at time $t=2$.  Therefore we impose:
\begin{equation}
\label{eq:gamma_equal}
\gamma_{1}^{Y_1(0)} = \gamma_{0}^{Y_1(1)} = \gamma_{0}^{Y_1(0)Y_1(1)}=\gamma_{1}^{Y_1(0)Y_1(1)} = 0
\end{equation}
assuming that
\begin{eqnarray}\label{eq:W2_SI}
W_{i2} = 1 & \hbox{if} &  W_{i2}^\ast \equiv \gamma_{w_1} + \gamma_{w_1}^{Y_1(w_1)} Y_{i1}(w_1)  + \epsilon_{i,W_2} > 0 
\end{eqnarray}
where $\epsilon_{i,W_2} \sim N(0,1)$, $w_1 \in \{0,1\}$,  $Y_{i1}(w_1) \in  \{0, 1 \}$.
 Thus, under SI/SIL we have:
$$
h_{iy_1}^{w_1} = 
\begin{cases}
\Phi \left( \gamma_{w_1} \right) & \hbox{ if } W_{i1}=w_1 \hbox { and }  y_1=0,\\
\Phi \left( \gamma_{w_1} + \gamma_{w_1}^{Y_1(w_1)}  \right) & \hbox{ if } W_{i1}=w_1 \hbox { and }  y_1=1. \\
\end{cases}
$$
with $w_1 \in \{0,1\}$. 

Finally, we need a model for the final outcome, $Y_{i2}$. In the empirical example  the final outcome, $Y_{i2}$, is the number of employees, which we consider as a continuous variable. Consistently we focus on a continuous final outcome in the simulation studies. 
Specifically we specify  Normal distributions for $Y_{i2}$ conditional on principal strata:
\begin{eqnarray}\label{eq:Y2}
\lefteqn{Y_{i2}(w_1, w_2)| G_{i}=g   \sim }\\&&  N \Bigl( \beta_{w_1w_2} + \beta_{w_1w_2}^{Y_1(0)} Y_{i1}(0) + \beta_{w_1w_2}^{Y_1(1)} Y_{i1}(1) + \beta_{w_1w_2}^{Y_1(0)Y_1(1)} Y_{i1}(0) Y_{i1}(1); \, \sigma_{w_1w_2,g}^2 \Bigr),\nonumber
\end{eqnarray}
$w_1\in \{0,1\}$, $w_2\in \{0,1\}$, $g\in \{00,01,10,11\}$.
For simplicity, we impose prior equality of the variance parameters across principal strata: $\sigma_{w_1w_2, g}^2= \sigma_{w_1w_2}^2$, $g\in \{00,01,10,11\}$. 
Then, the Normal distributions in  Equation~\eqref{eq:Y2} implies that:
$$
f_{ig}^{w_1,w_2}(y_2) = 
\begin{cases}
f \Bigl(y_2; \beta_{w_1w_2}, \, \sigma_{w_1w_2}^2 \Bigr) & \hbox{ if }  W_{i1}=w_1, W_{i2}=w_2,  \hbox { and }  G_i=00;\\
 f \Bigl(y_2; \beta_{w_1w_2}  + \beta_{w_1w_2}^{Y_1(0)}, \, \sigma_{w_1w_2}^2 \Bigr) & \hbox{ if } W_{i1}=w_1, W_{i2}=w_2,  \hbox { and }  G_i=10;\\
 f \Bigl(y_2; \beta_{w_1w_2}  + \beta_{w_1w_2}^{Y_1(1)}, \, \sigma_{w_1w_2}^2 \Bigr) & \hbox{ if }  W_{i1}=w_1, W_{i2}=w_2,   \hbox { and } G_i=01;\\
 f \Bigl(y_2; \beta_{w_1w_2} + \beta_{w_1w_2}^{Y_1(0)}  + \beta_{w_1w_2}^{Y_1(1)}  + \beta_{w_1w_2}^{Y_1(0)Y_1(1)}, \, \sigma_{w_1w_2}^2 \Bigr) & \hbox{ if }  W_{i1}=w_1, W_{i2}=w_2,   \hbox { and } G_i=11;\\
\end{cases}
$$
$w_1 \in \{0,1\}$, $w_2 \in \{0,1\}$, where $f(\cdot; \mu, \sigma^2)$ is the probability density function of a normal distribution with mean $\mu$ and variance $\sigma^2$. We use this model specification to generate data for the final outcome under both LSI and SI/SIL.
%In the first two simulation scenarios where there is no unmeasured confounder, and thus SI/SIL and LSI respectively hold without conditioning on $U_i$, data are generated setting $\alpha^{11}_U=\alpha^{00}_U=\alpha^{10}_U=0$, $\gamma_U=0$ and $\beta_U=0$.

The complete parameter vector for the simulation models is $\btheta=(\balpha, \bgamma, \bbeta, \bsigma^2)$, where
$\balpha=(\alpha^{11},   \alpha^{00},  \alpha^{10})$, $\quad \bgamma= (\{\gamma_{w_1}, \gamma_{w_1}^{Y_1(0)}, \gamma_{w_1}^{Y_1(1)}, \gamma_{w_1}^{Y_1(0)Y_1(1)}\}_{w_1 \in \{0,1\}})$, $\quad \bbeta=(\{\beta_{w_1w_2}, \beta_{w_1w_2}^{Y_1(0)}, \beta_{w_1w_2}^{Y_1(1)}, \beta_{w_1w_2}^{Y_1(0)Y_1(1)}\}_{w_1 \in \{0,1\},w_2 \in \{0,1\}})\,$ and $\bsigma^2=$ $\{\sigma_{w_1w_2}^2\}_{w_1 \in \{0,1\},w_2 \in \{0,1\}}$.  The parameter vector  $\btheta$ includes 31 parameters, 4 of which are forced to be equal to 0 when SI/SIL holds according to Equation \eqref{eq:gamma_equal}. %and 5 of which are forced to be equal to 0 in absence of unmeasured confounding.
The true values of all the parameters are given in the Supplementary Material available on-line.

%We first focus on the two scenarios where data are generated assuming that no unmeasured covariate enter either the assignment mechanism or the model for $Y_{i2}$, that is, imposing $\alpha_U^{11} =\alpha_U^{00}=\alpha_U^{10}=0$, $\gamma_U=0$ and $\beta_U=0$. 
%We then focus on the $2 \times 4$ sub-scenarios where data are generated assuming that there exist a confounder $U_i$ of the relationship between $W_{i2}$ and  $Y_{i2}$ that is associated with $G_i$. These sub-scenarios are defined by varying the strength of association of $U_i$ with $W_{i2}$ and  $Y_{i2}$, and the strength of association of $U_i$ with $G_i$, which depend on the parameters $\gamma_U$ and $\beta_U$, and  $\alpha_U^{11}, \alpha_U^{00}$ and $\alpha_U^{10}$, respectively (see Table \ref{tab:U}).  
%\begin{table}
%	\caption{\label{tab:U} True values for the parameters governing the association of $U_i$ with $W_{i2}$,  $Y_{i2}$, and $G_i$.}
%	\centering
%	\begin{tabular}{l|cc}
%		Association of $U_i$  & \multicolumn{2}{c}{Association of $U_i$ with $G_i$}\\
%		with $W_{i2}$ and $Y_{i2}$    & Weak & Strong \\
%		\hline
%		Weak   &  $\gamma_U=0.25$, $\beta_U	=0.25$ & $\gamma_U=0.25$, $\beta_U=0.25$\\
%		& $\alpha_U^{11}=0.20, \alpha_U^{00}=0.10, \alpha_U^{10}=0.02$ & $\alpha_U^{11}=0.45, \alpha_U^{00}=0.25, \alpha_U^{10}=0.15$\\
%		\\
%		Strong &   $\gamma_U=0.60$, $\beta_U=1$ & $\gamma_U=0.60$, $\beta_U=1$\\
%		& $\alpha_U^{11}=0.20, \alpha_U^{00}=0.10, \alpha_U^{10}=0.02$  & $\alpha_U^{11}=0.45, \alpha_U^{00}=0.25, \alpha_U^{10}=0.15$\\
%		\hline
%	\end{tabular}
%\end{table} 
%

\subsection{Inference in the simulations}
\label{sebsec:sim_inference}
For each simulation scenario we conduct inference under both LSI and SI/SIL.
Under LSI we assume the parametric models specified in Equations ~\eqref{eq:Y1}, ~\eqref{eq:W2_LSI} and ~\eqref{eq:Y2} in Section~\ref{subsec:mod}. It is worth noting that inference under LSI always uses a  correct model specification, even when SI/SIL holds.  In fact, when SI/SIL holds,
some parameters, namely $\gamma_{1}^{Y_1(0)}$, $\gamma_{0}^{Y_1(1)}$,  $\gamma_{0}^{Y_1(0)Y_1(1)}$ and $ \gamma_{1}^{Y_1(0)Y_1(1)}$, are simply equal to zero. 

Under the specification of type SI-1 we assume the parametric models specified in Equations ~\eqref{eq:Y1}, ~\eqref{eq:W2_SI} and ~\eqref{eq:Y2} in Section~\ref{subsec:mod}.
Under the specification of type SI-2 we specify the probit model in Equation~\eqref{eq:W2_SI} for treatment assignment  at time $t=2$, and the following models for the intermediate and final outcomes.  
We  use a probit specification for the intermediate outcome, $Y_{i1}(w_1)$:
\begin{eqnarray}\label{eq:Y1_SI}
Y_{i1}(w_1) = 1 & \hbox{if} &  Y_{i1}^\ast(w_1) \equiv \alpha_{w_1} + \epsilon_{i,Y_1(w_1)} > 0 
\end{eqnarray}
where $\epsilon_{i,Y_1(w_1)} \sim N(0,1)$, $w_1 \in \{0,1\}$, and we posit a Normal model on the final outcome 
$Y_{i2}(w_1, w_2)$, conditional on $Y_1(w_1)$:
\begin{equation}\label{eq:Y2_SI}
Y_{i2}(w_1, w_2)| Y_{i1}(w_1)  =y_1 \sim   N \Bigl( \beta_{w_1w_2} + \beta_{w_1w_2}^{Y_1(w_1)} y_1; \, \sigma_{w_1w_2,y_1}^2 \Bigr)
\end{equation}
$w_1\in \{0,1\}$, $w_2\in \{0,1\}$, $y_1\in \{0,1\}$. Again we impose  $\sigma_{w_1w_2, y_1=0}^2= \sigma_{w_1w_2, y_1=1}^2\equiv \sigma_{w_1w_2}^2$.
We assume that parameters are a priori independent and use proper, although weakly informative, prior distributions: in the initial setting we specify Normal priors with mean zero and variance 100  for the regression coefficients, and Scaled-Inverse-$\chi^2$ prior distributions with scale 1 and  0.002 degrees of freedom for the variance parameters. To asses the robustnes of the results to the specification of the prior distributions, different combinations of values for  the hyper-parameters (the variance of the Normal priors and the degree of freedom of the Scaled-Inverse-$\chi^2$ priors) are explored (see the Supplementary Material available on-line for details).

Posterior inference on $\btheta$ is obtained using Markov Chain Monte Carlo (MCMC) methods. 
The MCMC algorithms we adopted under LSI and SI/SIL with specification SI-1 use Gibbs sampler with data augmentation to impute at each step the missing principal stratum membership, $G_i$.  
Under SI/SIL with  specification SI-2, the likelihood function does not involve mixtures of distributions associated with the latent strata, but only depends on observed distributions, so the posterior distribution of $\btheta$ can be easily derived using Gibbs sampling methods. See on-line Supplementary Material for further details on the prior distributions and the MCMC algorithms.
The posterior distributions were simulated running a chain for 9000 MCMC iterations, after an initial 1000 burn-in iterations.   For  comparison purposes, when SI/SIL is assumed, we also draw inference on the causal effects of interest   using saturated marginal structural models, estimated by means of inverse probability of treatment weighting in a frequentist fashion. Results are shown in the Supplementary Material available on-line.

\subsection{Simulation Results}
\label{sec:resul}

\begin{figure}[ht]
\centering
\includegraphics[scale=0.55]{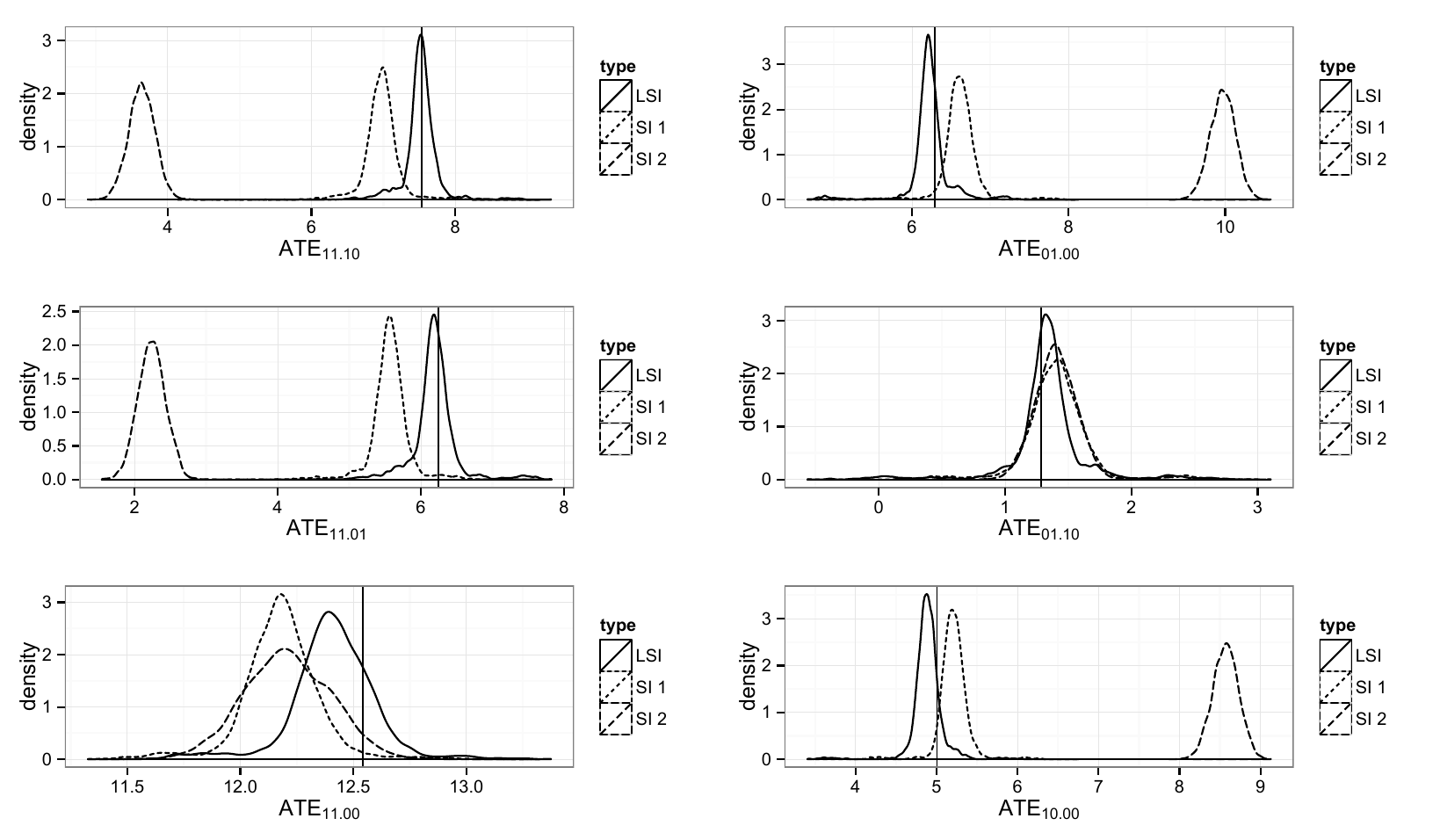}
\caption{\label{fig:LSI_eff}Posterior density functions for the ATEs under the LSI scenario. Inference using LSI (solid), SI-1 (dotted) and SI-2 (dashed). The vertical solid line indicates the true value of the ATE.} 
\end{figure}

\begin{figure}[ht]
\centering
\includegraphics[scale=0.55]{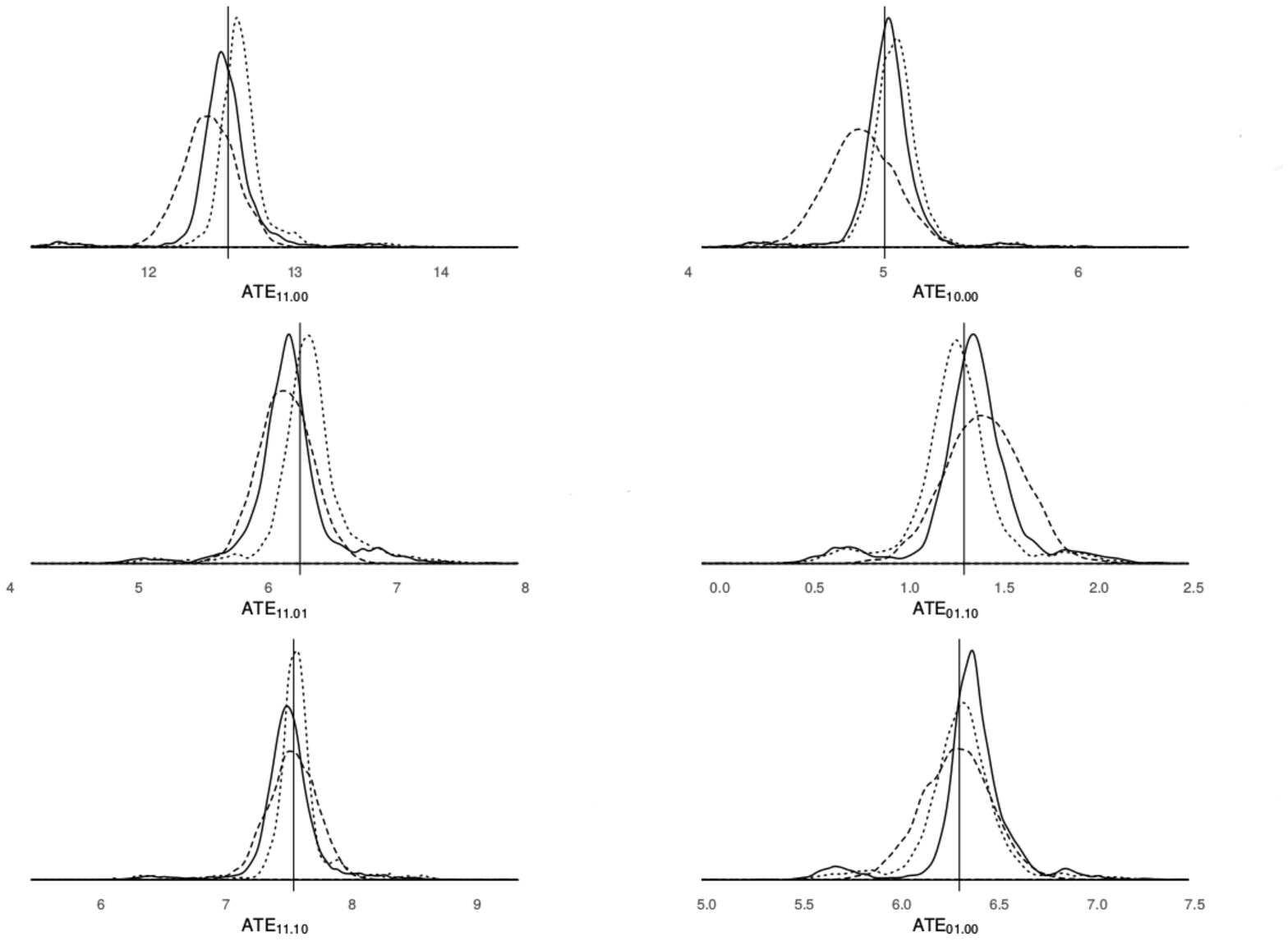}
\caption{\label{fig:SI_eff}Posterior density functions for the ATEs under the SI scenario. Inference using LSI (solid), SI-1 (dotted) and SI-2 (dashed).  The vertical solid line indicates the true value of the ATE.} 
\end{figure}

Simulation results for the causal estimands of interest are shown in Tables \ref{tab:ResLSI} and \ref{tab:ResSI} and in Figures  \ref{fig:LSI_eff} and \ref{fig:SI_eff}. Table \ref{tab:ResLSI} and Table \ref{tab:ResSI} show posterior means, standard deviations and 95\% posterior credible intervals for the average causal effects in Equation~\eqref{eq:ATE}  when LSI and SI/SIL holds, respectively, and inference is  conducted under LSI, and under SI/SIL using SI-1 and SI-2 specifications.  
Similarly, under the same scenarios and model specifications, 
Figures \ref{fig:LSI_eff} and \ref{fig:SI_eff} depict the posterior distributions of the six average causal effects.

Table \ref{tab:ResLSI} and  Figure \ref{fig:LSI_eff} make it clear that when LSI is the true assumption behind the data generating process, inference under sequential ignorability assumptions may lead to misleading results. 
Figure  \ref{fig:LSI_eff} shows that only the LSI inferential framework is able to always lead to valid inference about the six causal effects of interest. All the posterior distributions of the six $ATE$s derived under LSI 
reach their maximum in a thin neighbourhood around the true  $ATE$ values, so the posterior modes, which approximately correspond to the posterior means, appear to be  good point estimates for the causal effects of interest. Also the posterior variability is relatively small and the  95\% posterior credible intervals, which always cover the true $ATE$ values, are quite narrow, making inference 
very precise.
Conversely, assuming SI/SIL, when LSI actually holds, may yield to completely wrong inferences, especially when a specification of type SI-2 is used. 
The posterior distributions of the six $ATE$s derived under SI/SIL with a specification of type SI-1 cover the true $ATE$ values only in the queue for most of the six causal effects. Specifically, the $95\%$ posterior credible intervals cover the true $ATE$ values only for three out of the six $ATE$s: $ATE_{10.00}$, $ATE_{01.10}$, and $ATE_{01.00}$. For the remain causal estimands,  
$ATE_{11.00}$, $ATE_{11.01}$, and $ATE_{11.10}$, the true values are extreme values according to the estimated posterior distributions.
The handicaps of inference under  sequential ignorability assumptions is even more dramatic when a specification of type SI-2 is used.
In such a case, the posterior distributions for four out of six $ATE$s are located far away from the true values.

The different performances of inference under SI/SIL, comparing specifications of type SI-1 and SI-2, %(and MSM), 
may be (at least partially) justified noting that the first specification accounts for heterogeneity in the distribution of the final outcome, and thus of the causal effects, across principal strata. Therefore, although focus is on average causal effects for the whole population, if principal strata are strongly associated with the final outcomes, a parametrization of type SI-1 may, in some sense, address the consequences of a misspecified treatment assignment mechanism.

\begin{table}
\footnotesize
\caption{\label{tab:ResLSI} Summary statistics of the posterior distributions for the causal estimands when LSI holds.}
\centering
%\hspace*{-1.8cm}
\begin{tabular}{l c |c c c c | c c c c | c c c c }
\hline
&  & \multicolumn{4}{c|}{LSI} & \multicolumn{4}{c}{SI-1} & \multicolumn{4}{c}{SI-2}  \\
\cline{3-14} 
Estimand & true & mean & sd & $2.5\%$ & $97.5\%$ & mean & sd & $2.5\%$ & $97.5\%$ & mean & sd & $2.5\%$ & $97.5\%$ \\
\hline
$ATE_{11.00}$ & 12.54 & 12.41 & 0.19 & 11.91 & 12.76 & 12.17 & 0.17 & 11.74 & 12.52 & 12.21 & 0.19 & 11.84 & 12.58 \\
$ATE_{11.01}$ & 6.25 & 6.17 & 0.32 & 5.50 & 6.92 & 5.56 & 0.26 & 5.01 & 6.15 & 2.24 & 0.19 & 1.87 & 2.61 \\
$ATE_{11.10}$ & 7.54 & 7.52 & 0.25 & 6.96 & 7.99 & 6.97 & 0.23 & 6.43 & 7.46 & 3.64 & 0.18 & 3.27 & 4.00 \\
$ATE_{10.00}$ & 5.01 & 4.89 & 0.20 & 4.61 & 5.26 & 5.21 & 0.18 & 4.93 & 5.51 & 8.57 & 0.16 & 8.25 & 8.88 \\
$ATE_{01.10}$ & 1.29 & 1.33 & 0.32 & 0.44 & 2.14 & 1.41 & 0.28 & 0.79 & 2.05 & 1.40 & 0.16 & 1.09 & 1.71 \\
$ATE_{01.00}$ & 6.29 & 6.22 & 0.27 & 5.57 & 6.77 & 6.61 & 0.22 & 6.21 & 6.96 & 9.97 & 0.16 & 9.64 & 10.28 \\
\hline
\end{tabular}
\end{table}

When SI/SIL is the true assumption underlying the data generation process, both LSI and SI/SIL (with either a specification of type SI-1 or a specification of type SI-2) lead to valid inferences about the causal estimands of interest. 
As can be seen in Figure \ref{fig:SI_eff} and Table \ref{tab:ResSI}, the posterior distributions are  bell-shaped approximately symmetrical curves around the true $ATE$ values, providing relatively narrow 95\% posterior credible intervals, which always cover the true  $ATE$ values.

These results suggest that comparing inferences about the causal effects of interest derived under LSI and SI/SIL may provide useful insights on the plausibility of sequential ignorability assumptions. Further evidence can be obtained looking
at the posterior distributions of the treatment assignment probabilities at time $t=2$, given the treatment received at time $t=1$ and principal stratum membership, derived under the LSI assumption, and investigating if
equalities in Equation~\eqref{eq:equal} may hold. Equalities in Equation~\eqref{eq:equal} are indeed key quantities to assess the plausibility of sequential ignorability assumptions. 
LSI is a relaxed version of SIL: SIL can be viewed as a special case of LSI, where the equalities in Equation~\eqref{eq:equal} hold. Therefore if equalities in Equation~\eqref{eq:equal} do not hold, SIL clearly does not hold either.
  If equalities in Equation~\eqref{eq:equal} do not hold,  doubts on the plausibility 
of SI (Assumption \ref{ass:SI}) also arise, because we can reasonably expect that the assignment probabilities at time $t=2$ depend on unobserved characteristics related to the final outcome, which make SI untenable.
	 
\begin{table}
\footnotesize
\caption{\label{tab:ResSI} Summary statistics of the posterior distributions for the causal estimands when SI/SIL holds.}
\centering
% \hspace*{-1.8cm}
\begin{tabular}{l c |c c c c | c c c c | c c c c  }
\hline
&  & \multicolumn{4}{c|}{LSI} & \multicolumn{4}{c}{SI-1} & \multicolumn{4}{c}{SI-2}  \\
\cline{3-14} 
Estimand & true & mean & sd & $2.5\%$ & $97.5\%$ & mean & sd & $2.5\%$ & $97.5\%$ & mean & sd & $2.5\%$ & $97.5\%$ \\
\hline
$ATE_{11.00}$ & 12.54 & 12.52 & 0.25 & 12.00 & 12.98 & 12.62 & 0.26 & 12.13 & 13.08 & 12.42& 0.18 & 12.06 & 17.77\\
$ATE_{11.01}$ & 6.25  & 6.16 & 0.31 & 5.51 & 6.91 & 6.31 & 0.31 & 5.60 & 7.00 & 6.14 & 0.21 & 5.73 & 6.55 \\
$ATE_{11.10}$ & 7.54  & 7.49 & 0.27 & 6.85 & 8.12 & 7.56 & 0.28 & 6.88 & 8.22 & 7.53 & 0.19 & 7.14 & 7.90 \\
$ATE_{10.00}$ & 5.01  & 5.02 & 0.17 & 4.65 & 5.31 & 5.06 & 0.17 & 4.68 & 5.35 & 4.89 & 0.18 & 4.55 & 5.23 \\
$ATE_{01.10}$ & 1.29  & 1.33 & 0.25 & 0.62 & 1.90 & 1.24 & 0.24 & 0.66 & 1.85 & 1.39 & 0.21 & 0.99 & 1.79 \\
$ATE_{01.00}$ & 6.29  & 6.35 & 0.23 & 5.66 & 6.84 & 6.30 & 0.21 & 5.78 & 6.80 & 6.28 & 0.18 & 5.92 & 6.63 \\
\hline
\end{tabular}
\end{table}

\begin{figure}[ht]
\centering
\includegraphics[scale=0.55]{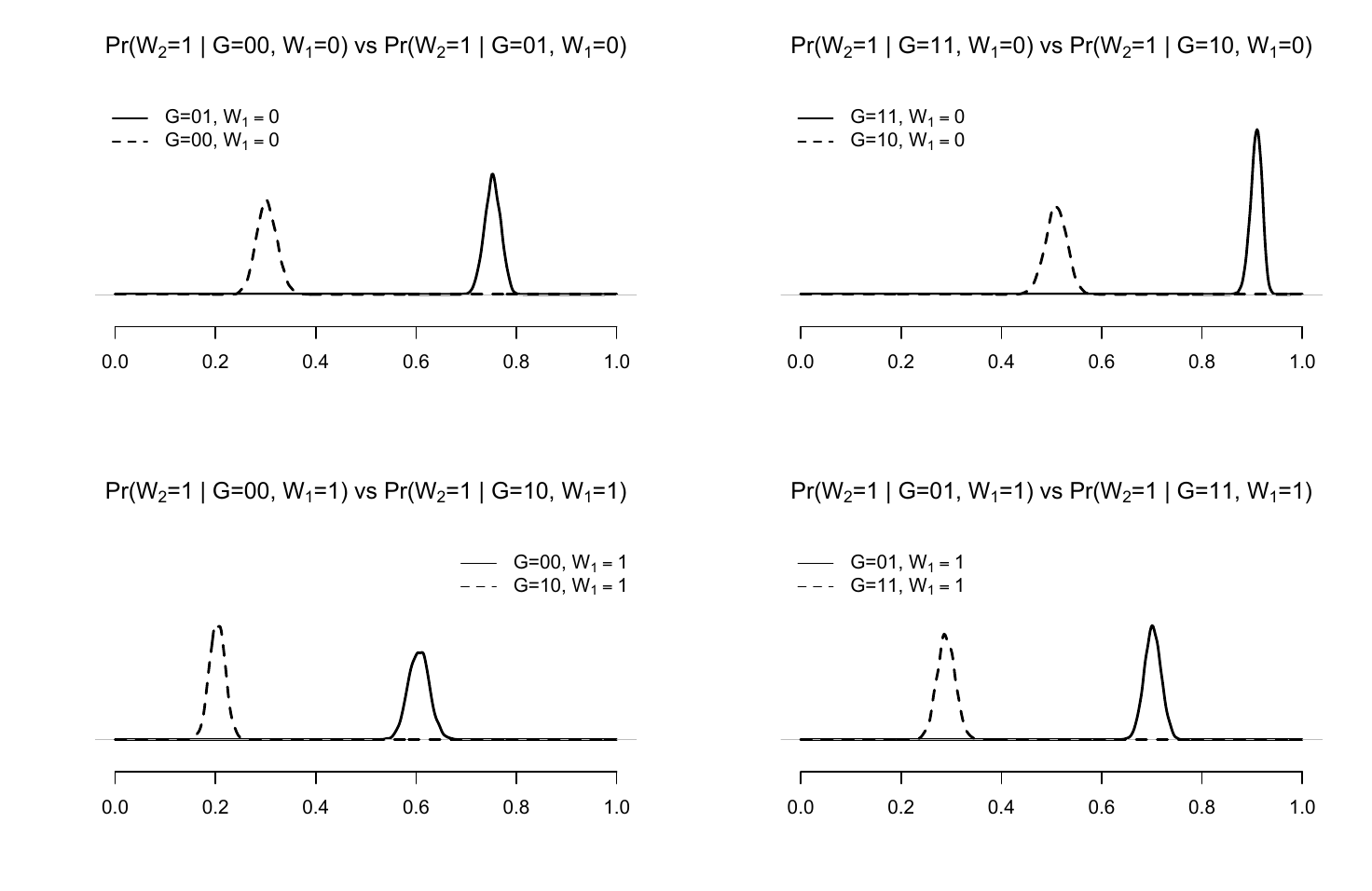}
\caption{\label{fig:LSI_W2}Posterior density functions for the assignment probabilities at time $t=2$ when LSI holds.} 
\end{figure}
\begin{figure}[ht]
\centering
\includegraphics[scale=0.55]{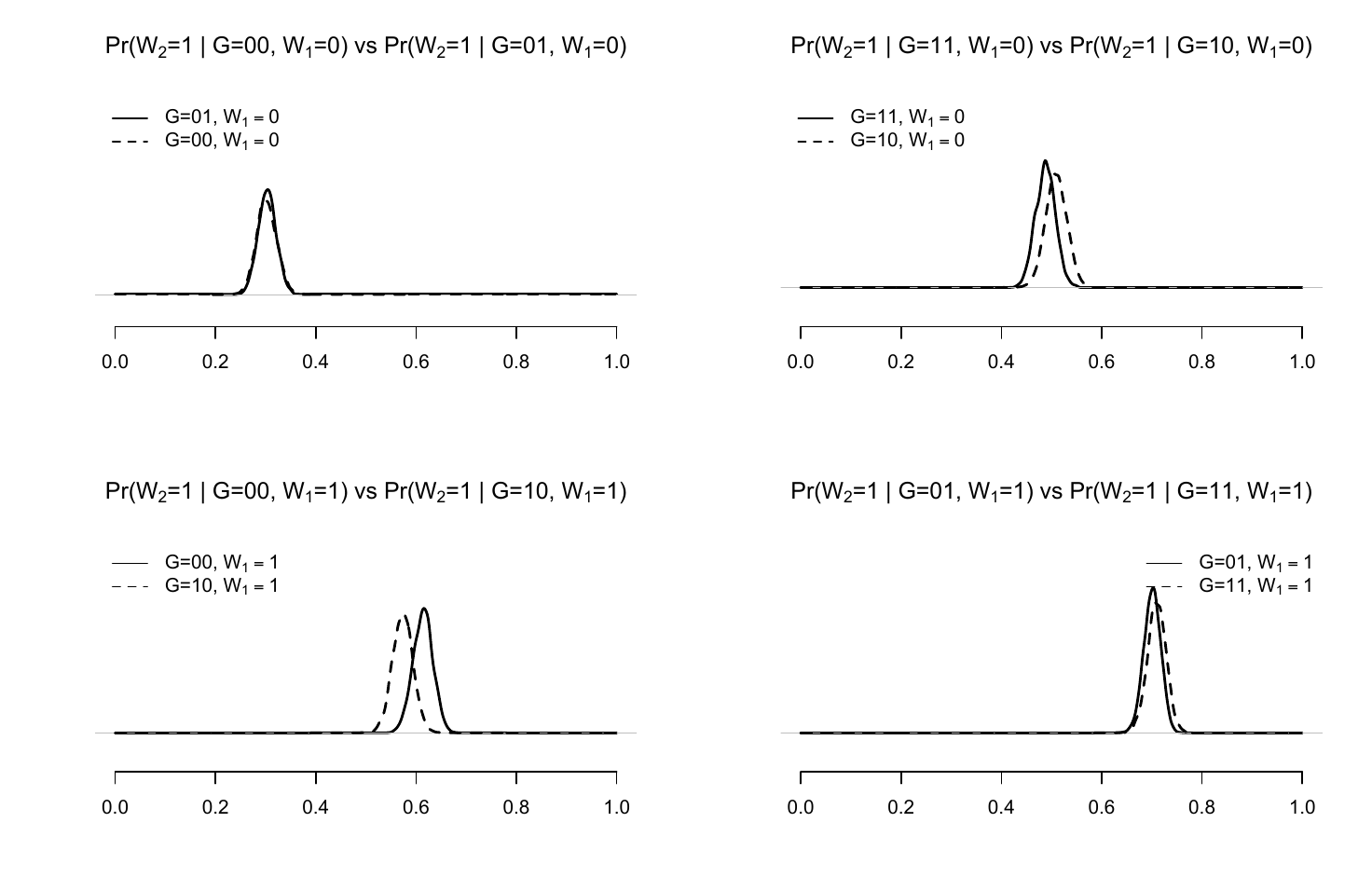}
\caption{\label{fig:SI_W2}Posterior density functions for the assignment probabilities at time $t=2$ when SI holds.} 
\end{figure}
Figures  \ref{fig:LSI_W2} and \ref{fig:SI_W2} show the  posterior distributions of those treatment assignment probabilities     derived when LSI holds and SI/SIL holds, respectively. The posterior distributions are coupled according to the equalities in Equation~\eqref{eq:equal}.
As we could expect, the posterior distributions of the treatment assignment probabilities at time $t=2$  in Figure \ref{fig:LSI_W2} appear to be highly heterogeneous across principal strata, suggesting strong evidence against the equalities in Equation~\eqref{eq:equal}, and thus, against sequential ignorability. 
On the other hand, the posterior distributions  in Figure \ref{fig:SI_W2}  look very similar and are highly  overlapping, showing no evidence against sequential ignorability in favor of LSI.

We investigate the robustness of these results with respect to the specification of prior distributions using both more
informative priors  as well as  less informative priors.
The results appear to be robust with respect to the specification of the prior distributions: different prior specifications change the results only slightly without affecting the substantive conclusions (see 
% Table~\ref{tab:SensLSI} and Figures~\ref{fig:SensLSIa}-\ref{fig:SensLSIc} and
%Table~\ref{tab:SensSI} and Figures~\ref{fig:SensSIa}-\ref{fig:SensSIc} 
Table S5 and Figures S1-S3 and
Table S6 and Figures S4-S6
in the Supplementary Material available on-line).

In the Supplementary Material available on-line (Tables S8-S9 and Figures S7-S10) we also investigate and compare the role of the alternative assumptions  (SI/SIL and LSI) when they hold conditional on an  unmeasured confounder, which can be (partially) explained by latent principal stratum membership, $G_i$. 
Specifically we consider two additional simulation scenarios  where data are generated under SI/SIL and LSI, respectively, conditioning on a binary covariate,   $U_i$, such that $(1)$ it is  related to both treatment assignment at time $t=2$, $W_{i2}$, and the final outcome, $Y_{i2}$; and $(2)$ it is associated with principal stratum membership, $G_i$. 
The two scenarios  are defined by varying   the strength of association of $U_i$ with $G_i$. 
We consider $U_i$  as an unmeasured confounder, and thus we conduct inference by ignoring it.
The key result we obtain is that the posterior distribution of the   assignment probabilities at time $t=2$ derived under LSI show some evidence against SI/SIL in any scenario we consider, even when data are generated under SI/SIL conditional on the unobserved confounder, $U_i$. Consistently inference under LSI generally performs better.%(see the bottom panel of  Figures \ref{fig:LSIUw},   \ref{fig:LSIUs}, \ref{fig:SIUw}, and \ref{fig:SIUs}). 

It is worth noting that our simulation results are based on a single simulated data set. Studying the frequentist properties  of our procedure in  repeated samples is beyond the scope of the paper. 

\section{An illustrative example: the Tuscan Government Founding Program}
\label{sec:Case}
We illustrate our framework in a program evaluation study concerning causal effects on firms' performances of an interest free loans policy aiming to ease access to credit by making it less costly.
Firms meeting certain standards to be eligible can apply to get an interest free loan at various points in time, thus firms may apply and be granted multiple times over subsequent years.

The program started in 2002 and was rolling on a yearly basis. In this paper we consider data in the years between 2002 and 2007, and we focus on casual effects defined by contrasting firms' performances measured in terms of employment levels at the end of the study, in $2007$, under different treatment sequences. Treatment sequences are defined using two binary treatment variables: $W_{i1}$ equal to one if firm $i$ is granted  at least one time between 2002 and 2004, and zero otherwise; and $W_{i2}$ equal to one  if firm $i$ is granted at least one time between 2005 and 2007, and zero otherwise. 
The final outcome of interest, $Y_{i2}$, is the firm $i$'s number of employees at the end of 2007. As intermediate outcome we consider a binary variable $Y_{i1}$ describing the hiring policy of firm $i$ at the end of 2004: $Y_{i1}=1$ if firm $i$ hires new employees by the end of 2004, and $Y_{i1}=0$ otherwise.
Therefore the basic principal stratification with respect to this intermediate variable classifies firms into four latent groups: $G_i=00$ comprising firms that would not hire irrespective of the treatment received at time $t=1$; $G_i=01$ comprising firms that would hire if granted but would not hire if  not granted; $G_{i} = 10$ comprising firms that would hire if not granted and would hire if granted; and   $G_i=11$ comprising firms that would always hire irrespective of the treatment received at time $t=1$.
Principal strata with respect to the indicator for hiring choices can be viewed as a coarsened representation of the latent hiring preferences of a firm, which are reasonably associated with both the decision to participate in the treatment at a subsequent time point and  the final outcome.

Firms exposed to different treatment sequences are likely to differ in many dimensions. Thus, preliminary analyses were conducted to create sub-populations of firms, 
where the distributions of the baseline background variables are well balanced across firms that received at least a loan and firms that did not. 
Specifically the data set we have comprises a sample of firms, which was selected  using the following matching procedure.
First all firms that received a loan in at least one treatment period (treated firms) were included in the sample.
Then a sub-sample of firms that did not receive any loan in the observation period (control firms)  was selected   matching each treated firm to $k=6$  control firms using nearest-neighbor propensity score matching, where the propensity score is defined as the conditional probability of receiving a loan in at least one treatment period given the baseline background characteristics  (see  Pirani et al. 2013 for details). 
From this sample we delete firms that cease their operations during the observation period.

Our final sample consists of $4\,615$ firms, among which 632 firms received a loan only in one treatment period,  33 firms received a loan in both treatment periods, and the rest did not received any loan in the observation period.	Before moving to the analysis phase, we conducted additional preliminary analyses. We checked that the distribution of the baseline observed covariates were	well balanced with respect to the treatment assigned at time $t=1$, $W_{i1}$, that is, between firms that received a loan at time $t=1$  and firms that did not receive a loan at time $t=1$. 	The balance in the covariate distributions appears to be good in terms of normalized mean differences \cite[e.g.,][]{ImbensRubin:2015}: across the pre-treatment baseline variables, the maximum value of the normalized difference in covariate means is $0.35$ and the normalized difference  is less than 0.25 for most of the covariates. 	This result suggests that in our sample of firms 	the degree of balance with respect to $W_{i1}$ is comparable to what one might expect in a completely randomized experiment. 
Therefore we can reasonably analyze the sample data as coming from a (quasi-) randomized experiment, and assume that the treatment at time $t = 1$ is randomly assigned. We then assume that treatment at time $t = 2$ is randomly assigned conditional on the observed value of the intermediate outcome under SI/SIL, and conditional on principal stratum membership under LSI\footnote{We also conducted our Bayesian analyses conditioning on covariates. Results change only slightly; the presence of covariates mainly affects the posterior variability of the causal estimands, introducing noise. Therefore  we preferred to focus on results derived without conditioning on covariates, also in line with our simulation study.}.

Bayesian inference for the average casual effects of interest was conducted under both LSI and SI/SIL. Under LSI the model we specified involved the tree sub-models described in Equations \eqref{eq:Y1}, \eqref{eq:W2_LSI} and \eqref{eq:Y2}. Bayesian inference under SI/SIL was conducted using both a specification of type SI-1, which involved the tree sub-models described in Equations \eqref{eq:Y1}, \eqref{eq:W2_SI} and \eqref{eq:Y2}, and a specification of type SI-2, which involved the tree sub-models described in Equations \eqref{eq:Y1_SI}, \eqref{eq:W2_SI} and \eqref{eq:Y2_SI}. 

\begin{figure}[ht]
\centering
\includegraphics[scale=0.6]{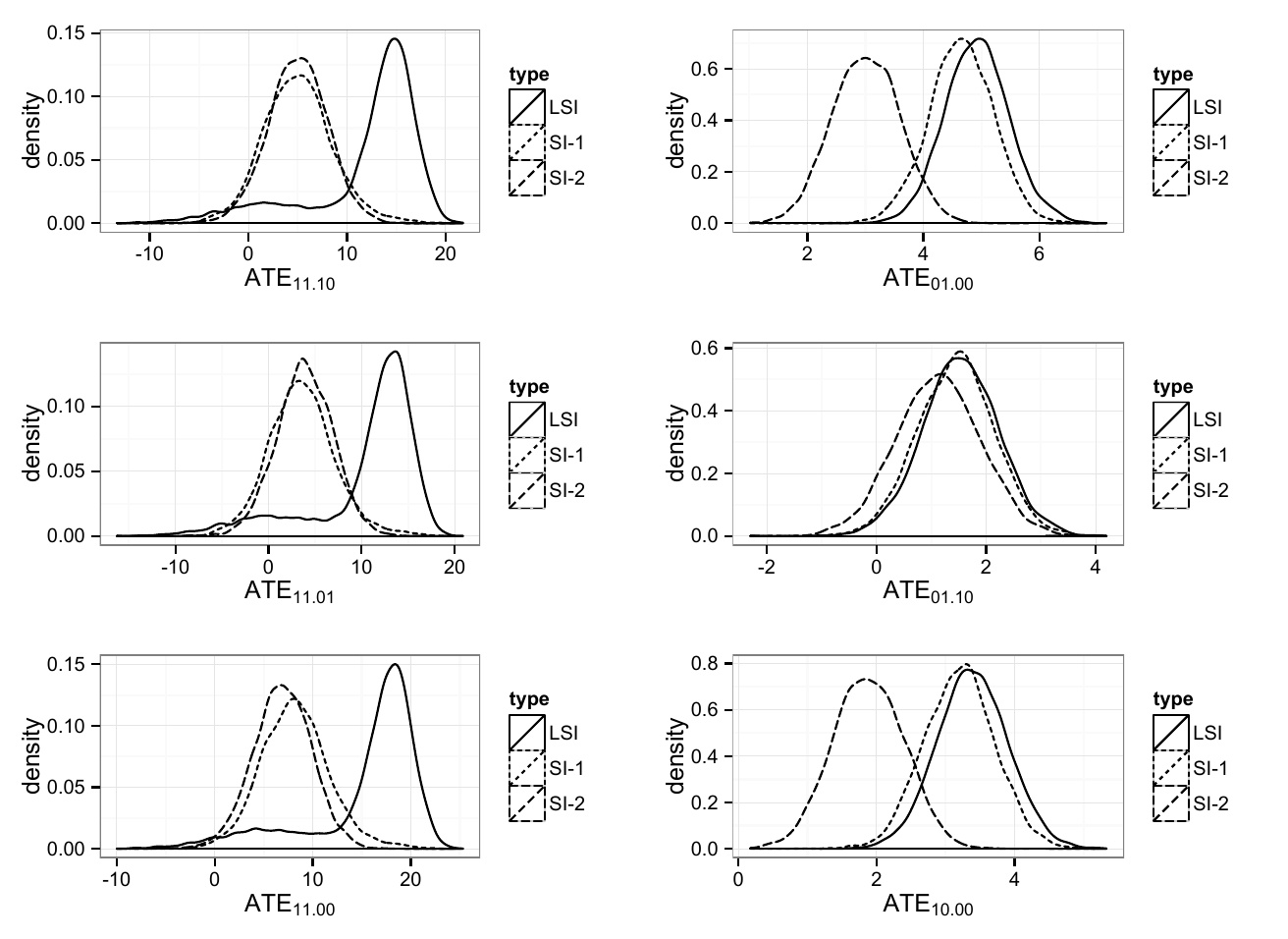}
\caption{\label{fig:REAL_eff_3}Posterior density functions for the ATEs in the real case application. Inference using LSI (solid), SI-1 (dotted) and SI-2 (dashed).} 
\end{figure}

Table \ref{tab:ResREAL_3} shows the posterior means, standard deviations and 95\% posterior credible intervals for the stratum membership probabilities and the six $ATE$s, while Figure \ref{fig:REAL_eff_3}  portrays the posterior density functions of the six $ATE$. 
%Table \ref{tab:ResREAL_3} also shows means, standard errors and 95\%  confidence obtained using MSMs.
Inference under SI/SIL does not seem to strongly depend on the type of specification used. 
Specifications SI-1 and SI-2 % and analysis based on MSM 
lead to similar results for all the six causal effects of interest but two, $ATE_{01.00}$ and $ATE_{10.00}$. 
The  posterior means   of  $ATE_{01.00}$ and $ATE_{10.00}$ derived under specification SI-1 are 
greater than those derived under specification SI-2.% and than the MSM-based estimates. 
%Finally, when MSM is used to obtain $ATE_{01.10}$, the mean effect is considerably lower than those obtained under all the other modelling approaches. 

Under LSI we obtain substantially different inferential results than those under SI/SIL, especially for some causal estimands.
  LSI leads to posterior distributions for the causal effects  $ATE_{10.00}$, $ATE_{01.10}$ and $ATE_{01.00}$  that are
	essentially the same as those derived under SI/SIL with a specification of type SI-1.
	However, LSI and SI/SIL provide very different posterior distributions for $ATE_{11.00}$, $ATE_{11.01}$ and $ATE_{11.10}$. %irrespective of the type of specification we use under SI/SIL.  
	Under LSI we obtain posterior distributions for $ATE_{11.00}$, $ATE_{11.01}$ and $ATE_{11.10}$  that are centered on much higher values and have a higher variability than those derived under SI/SIL, irrespective of the type of specification we use under SI/SIL.
Specifically the posterior means of $ATE_{11.00}$, $ATE_{11.01}$ and $ATE_{11.10}$  derived under LSI are more than 1.8 times those derived under SI/SIL, but the posterior standard deviations  of $ATE_{11.00}$, $ATE_{11.01}$ and $ATE_{11.10}$ derived under LSI are about twice those derived under SI/SIL. 

The loss of precision we have assuming LSI rather than SI/SIL  is probably due to the greater complexity of the model, which is only weakly identified \citep{Gustafson:2010}; the small proportion of firms that is estimated to belong to some principal strata (e.g., the posterior means of the probabilities to belongs to groups $01$ and $11$ are equal to $0.04$, see $ {\pi}_g$ in Table~\ref{tab:ResREAL_3}); and the extremely low estimated probability to receive
a loan at time $t=2$ for all firms but those in the principal stratum $01$ that received a loan at time $t=1$ (see Table~\ref{tab:PrW2_REAL_3}).
\begin{table}
\footnotesize
\caption{\label{tab:ResREAL_3} Summary statistics of the posterior distributions for the causal estimands in the real case.}
\centering
% \hspace*{-1.5cm}
\begin{tabular}{l  |c c c c | c c c c | c c c c}
\hline
&   \multicolumn{4}{c|}{LSI} & \multicolumn{4}{c}{SI-1} & \multicolumn{4}{c}{SI-2}  \\
\cline{2-13} 
Estimand 		& mean & sd & 2.5\% & 97.5\% & mean & sd & 2.5\% & 97.5\% & mean & sd & 2.5\% & 97.5\% \\
   \hline
$\pi_{00}$ & 0.69 & 0.01 & 0.67 & 0.70 & 0.69 & 0.01 & 0.67 & 0.70 & $-$ & $-$ & $-$ & $-$  \\ 
$\pi_{01}$ & 0.04 & 0.00 & 0.03 & 0.05 & 0.04 & 0.00 & 0.03 & 0.04 & $-$ & $-$ & $-$ & $-$  \\ 
$\pi_{10}$ & 0.23 & 0.01 & 0.22 & 0.25 & 0.24 & 0.01 & 0.22 & 0.25 & $-$ & $-$ & $-$ & $-$  \\ 
$\pi_{11}$ & 0.04 & 0.00 & 0.03 & 0.05 & 0.04 & 0.00 & 0.03 & 0.05 & $-$ & $-$ & $-$ & $-$  \\    
$ATE_{11.00}$ & 15.11 & 6.03 & -0.69 & 21.62 & 8.22 & 3.47 & 1.63 & 15.70 & 6.88 & 2.96 & 0.90 & 12.66 \\ 
$ATE_{11.01}$ & 10.19 & 6.07 & -5.69 & 16.83 & 3.57 & 3.53 & -3.15 & 11.20 & 3.87 & 3.01 & -2.16 & 9.58 \\ 
$ATE_{11.10}$ & 11.73 & 6.01 & -3.97 & 18.36 & 5.02 & 3.50 & -1.60 & 12.52 & 5.00 & 3.00 & -1.06 & 10.85 \\ 
$ATE_{10.00}$ & 3.39 & 0.51 & 2.38 & 4.39 & 3.20 & 0.51 & 2.21 & 4.18 & 1.88 & 0.52 & 0.89 & 2.88  \\ 
$ATE_{01.10}$ & 1.54 & 0.70 & 0.15 & 2.93 & 1.45 & 0.70 & 0.09 & 2.80 & 1.13 & 0.77 & -0.39 & 2.59 \\ 
$ATE_{01.00}$ & 4.92 & 0.55 & 3.85 & 6.07 & 4.65 & 0.56 & 3.52 & 5.71 & 3.01 & 0.58 & 1.90 & 4.13\\ 
   \hline
\end{tabular}
\end{table}
 
Given that LSI and SI/SIL lead to quite different posterior distributions of the causal estimands of interest, it becomes of compelling interest to investigate the plausibility of the sequential ignorability assumptions. To this end, we look at the posterior distributions of the assignment probabilities at time $t=2$ derived under LSI,  summarized in Table \ref{tab:PrW2_REAL_3}.
As we can see,  
 the probability to receive a loan in the second treatment period is very small for all firms except for those belonging to principal stratum 01 that received a loan in the first treatment period ($W_{i1}=1$). 
Also we find large differences comparing the posterior distributions of the assignment probabilities at time $t=2$ two by two according to the equalities in Equation \eqref{eq:equal}. 
Among firms that did not receive a loan in the first treatment period ($W_{i1}=0$) the posterior mean of the probability  to receive a loan in the second treatment period is about $7\%$ for firms in principal stratum $10$ and about $1\%$ for firms in principal stratum 11. Among firms that received a loan in the first treatment period ($W_{i1}=1$) the posterior mean of the probability to receive a loan in the second treatment period is about 76\% for firms in principal stratum 01 and less than $0.5\%$ for firms in principal stratum 11.
The posterior distributions of the probability  to receive a loan in the second treatment period for firms in  principal strata 00 and 01 assigned to  $W_{i1}=0$ and for firms in  principal strata 00 and 10 assigned to  $W_{i1}=1$ are more similar, although  
there is still strong evidence against the assumption that they are the same, as they should be under SIL.
Therefore  our results show some evidence that sequential ignorability assumptions are questionable in this study, suggesting inference under LSI to be more reliable here. 
\begin{table}
\footnotesize
\caption{\label{tab:PrW2_REAL_3} Summary statistics for the assignment probabilities at time $t=2$ in the real case.}
\centering
\begin{tabular}{r r c c c c c}
\hline
$W_1$ & $G$ & mean & sd & $Q_1$ & median & $Q_3$ \\ 
\hline
& 00   & 0.0690 & 0.0048 & 0.0657 & 0.0689 & 0.0721 \\  
& 01   & 0.0432 & 0.0170 & 0.0309 & 0.0414 & 0.0532 \\ 
\cline{2-7} 
0 	& 10   & 0.0708 & 0.0081 & 0.0652 & 0.0706 & 0.0762 \\ 
& 11   & 0.0105 & 0.0100 & 0.0032 & 0.0074 & 0.0149 \\  
\hline  
& 00 	& 0.0890 & 0.0418 & 0.0795 & 0.0990 & 0.1156 \\
& 10    & 0.0765 & 0.1167 & 0.0139 & 0.0267 & 0.0538 \\ 
\cline{2-7} 
1	& 01   & 0.7627 & 0.1326 & 0.6785 & 0.7812 & 0.8643 \\ 
& 11   & 0.0004 & 0.0018 & 0.0000 & 0.0000 & 0.0000 \\
\hline
\end{tabular}
\end{table}

\section{Concluding Remarks}
\label{sec:conclusion}
We focus on the role of the critical assumptions about the assignment mechanism in causal inference for time-varying treatments, proposing a new assumption, that we call latent sequential ignorability (LSI), which may be more reasonable than  the usually invoked sequential ignorability assumptions in some settings. LSI implies that the joint values of potential outcomes for the relevant intermediate variables (i.e., the principal strata), rather than their observed values only, include crucial information about the decision to participate in the treatment. Therefore LSI focuses on specific violations of SI due to the presence of unobserved factors affecting the decision to participate in the treatment that can be summarized by  principal strata. 

In studies where ignorability assumptions are not reasonable, LSI provides a powerful framework, which also permits to easily assess the sensitivity of inferential conclusions with respect to violations of sequential ignorability assumptions (SI and SIL) implied by LSI looking at inferences on the probabilities of treatment assignment at a given time point under LSI, conditional on the observed history and principal strata. These quantities are key estimands in causal inference under LSI, so no additional effort is required to perform sensitivity analysis under LSI. 
 
Simulation results show that LSI conducts to valid inference for causal effects even if SI/SIL holds, although it usually involves more complex models. On the other hand, inference under a sequential ignorability assumption may lead to very misleading inferential conclusions when it does not hold, but LSI does. 

In our illustrative example, sensitivity analysis showed strong evidence against sequential ignorability assumptions: the posterior distributions of the treatment assignment probabilities at time $t=2$ within groups defined by the first treatment and principal stratum membership are quite heterogeneous, suggesting that inferences based on LSI are more reliable.

Another appealing feature of LSI is that it provides a natural framework to investigate the heterogeneity of the effects across principal strata. Assessing causal effects stratified by intermediate outcomes under SI/SIL generally requires additional efforts. 
In particular, in a Bayesian setting, one needs to specify a model for principal strata membership conditional on the covariates, and a model for the  potential outcomes $Y_{i2}(w_1,w_2)$ conditional on principal strata and covariates. This model specification, which corresponds to using  a specification of type SI-1 under SI/SIL, is the core of the inferential approach under LSI, but it is not  standard in causal inference under sequential ignorability assumptions. 
Here we did not investigate issues concerning causal effect heterogeneity across  principal strata, focusing on comparing inferences about causal effects for the whole population under LSI and sequential ignorability assumptions. Nevertheless the heterogeneity of the effects with respect to principal strata can be often of interest to policy makers.

As a general message, our study stresses the importance of carefully evaluating the plausibility of the assumptions underlying the analysis, especially in complex settings like those arising in longitudinal observational studies.

The extension of our framework to multiple (i.e., more than two) time points is without any doubt an interesting future development.
The extension to additional periods is conceptually straightforward, but raises challenging practical issues due to the
fact that the number of principal strata increases with the number of time points. To cope with the increasingly huge missing data problem, additional assumptions are required. For instance we could invoke  Markovian properties, similarly to what has been done, e.g., both by \citet[][]{Linetal:2008} in a study in which units are randomized at the baseline and compliance to treatment may vary longitudinally, and by \citet[][]{DaiGilbert:2012} who considered, again,  a single time treatment and some post-randomization time-varying behavioral variables.

\appendix
 
\section*{Appendix: Relationship between SIL and LSI}
\begin{proposition}
If Assumption \ref{ass:SI_AM} holds, then Assumption \ref{ass:LSI_AM} holds.
\end{proposition}
\medskip

\noindent \textbf{Proof.} The proof is articulated in three parts:
\begin{enumerate}
\item We first show that Assumption \ref{ass:SI_AM}  can be equivalently formulated 
using Equations \eqref{eq:SIL1} and \eqref{eq:SIL2}.
\item We then show that Assumption \ref{ass:LSI_AM} can be equivalently formulated 
using Equations \eqref{eq:LSI1} and \eqref{eq:LSI2}.
\item Finally we show that Equations \eqref{eq:SIL1} and \eqref{eq:SIL2} imply Equations \eqref{eq:LSI1} and \eqref{eq:LSI2}, and therefore Assumption \ref{ass:SI_AM} implies Assumption \ref{ass:LSI_AM}.
\end{enumerate}
\begin{description}
\item[\textit{$1.$}] \textit{Assumption \ref{ass:SI_AM} holds if and only if Equations \eqref{eq:SIL1} and \eqref{eq:SIL2} hold.}
\end{description}
\underline{Suppose that Assumption \ref{ass:SI_AM} holds}. By definition,
\begin{eqnarray*}
	\lefteqn{Pr(\bW_i \mid \bX_i , Y_{i1}(0), Y_{i1}(1), Y_{i2}(0, 0), Y_{i2}(1, 0), Y_{i2}(0, 1), Y_{i2}(1, 1))=}\\&&
	Pr(W_{i1} \mid \bX_i , Y_{i1}(0), Y_{i1}(1), Y_{i2}(0, 0), Y_{i2}(1, 0), Y_{i2}(0, 1), Y_{i2}(1, 1))
	\times \\&&
	Pr(W_{i2} \mid \bX_i, W_{i1},  Y_{i1}(0), Y_{i1}(1), Y_{i2}(0, 0), Y_{i2}(1, 0), Y_{i2}(0, 1), Y_{i2}(1, 1)).
\end{eqnarray*}
We have
\begin{eqnarray*} 
	\lefteqn{Pr(W_{i1} \mid \bX_i, Y_{i1}(0), Y_{i1}(1), Y_{i2}(0, 0), Y_{i2}(1, 0), Y_{i2}(0, 1), Y_{i2}(1, 1))=}\\&&
	\sum_{w_2=0,1}	Pr(W_{i1}, W_{i2}=w_2 \mid \bX_i, Y_{i1}(0), Y_{i1}(1), Y_{i2}(0, 0), Y_{i2}(1, 0), Y_{i2}(0, 1), Y_{i2}(1, 1))
	= \\&&
	\sum_{w_2=0,1}  Pr(W_{i1} \mid \bX_i) \times Pr(W_{i2}=w_2 \mid \bX_i, W_{i1}, Y_{i1}^{obs})
	=\\&&
	Pr(W_{i1} \mid \bX_i) \times  \sum_{w_2=0,1} Pr(W_{i2}=w_2 \mid \bX_i, W_{i1}, Y_{i1}^{obs})
	=  Pr(W_{i1} \mid \bX_i) 
\end{eqnarray*}
where the first equality follows from the law of total probability and  the second equality follows from Assumption \ref{ass:SI_AM}. Therefore Assumption \ref{ass:SI_AM} implies Equation \eqref{eq:SIL1}. Moreover 
\begin{eqnarray*}
	\lefteqn{Pr(W_{i2} \mid \bX_i, W_{i1},  Y_{i1}(0), Y_{i1}(1), Y_{i2}(0, 0), Y_{i2}(1, 0), Y_{i2}(0, 1), Y_{i2}(1, 1)) =}\\&&
	\dfrac{Pr(W_{i1}, W_{i2} \mid \bX_i, Y_{i1}(0), Y_{i1}(1), Y_{i2}(0, 0), Y_{i2}(1, 0), Y_{i2}(0, 1), Y_{i2}(1, 1))}{Pr(W_{i1} \mid \bX_i, Y_{i1}(0), Y_{i1}(1), Y_{i2}(0, 0), Y_{i2}(1, 0), Y_{i2}(0, 1), Y_{i2}(1, 1))}
	= \\&&
	\dfrac{Pr(W_{i1} \mid \bX_i) \times  P(W_{i2} \mid \bX_i,W_{i1}, Y_{i1}^{obs})}{ Pr(W_{i1} \mid \bX_i)}=
	P(W_{i2} \mid \bX_i,W_{i1}, Y_{i1}^{obs}) \\
\end{eqnarray*}
where the first equality holds by definition,  and the second equality follows from Assumption \ref{ass:SI_AM}. %and  the equality $Pr(W_{i1} \mid \bX_i, Y_{i1}(0), Y_{i1}(1), Y_{i2}(0, 0), Y_{i2}(1, 0), Y_{i2}(0, 1), Y_{i2}(1, 1))=Pr(W_{i1} \mid \bX_i) $ shown above. 

\underline{Vice-versa suppose that Equations \eqref{eq:SIL1} and \eqref{eq:SIL2} hold.} Then
\begin{eqnarray*}
	\lefteqn{Pr(\bW_i \mid \bX_i , Y_{i1}(0), Y_{i1}(1), Y_{i2}(0, 0), Y_{i2}(1, 0), Y_{i2}(0, 1), Y_{i2}(1, 1))=}\\&&
	Pr(W_{i1} \mid \bX_i , Y_{i1}(0), Y_{i1}(1), Y_{i2}(0, 0), Y_{i2}(1, 0), Y_{i2}(0, 1), Y_{i2}(1, 1))
	\times \\&&
	Pr(W_{i2} \mid \bX_i, W_{i1},  Y_{i1}(0), Y_{i1}(1), Y_{i2}(0, 0), Y_{i2}(1, 0), Y_{i2}(0, 1), Y_{i2}(1, 1))=\\&&
	Pr(W_{i1} \mid \bX_i)  \times  Pr(W_{i2} \mid W_{i1},  Y_{i1}^{obs}, \bX_i)
\end{eqnarray*}
where the first equality holds by definition and the second equality follows from Equations \eqref{eq:SIL1} and \eqref{eq:SIL2}.

\begin{description}
	\item[\textit{$2.$}] \textit{Assumption \ref{ass:LSI_AM} holds if and only if Equations \eqref{eq:LSI1} and \eqref{eq:LSI2} hold}
\end{description}
 \underline{Suppose that Assumption \ref{ass:LSI_AM} holds.} By definition,
 \begin{eqnarray*}
 	\lefteqn{Pr(\bW_i \mid \bX_i , Y_{i1}(0), Y_{i1}(1), Y_{i2}(0, 0), Y_{i2}(1, 0), Y_{i2}(0, 1), Y_{i2}(1, 1))=}\\&&
 	Pr(W_{i1} \mid \bX_i , Y_{i1}(0), Y_{i1}(1), Y_{i2}(0, 0), Y_{i2}(1, 0), Y_{i2}(0, 1), Y_{i2}(1, 1))
 	\times \\&&
 	Pr(W_{i2} \mid \bX_i, W_{i1},  Y_{i1}(0), Y_{i1}(1), Y_{i2}(0, 0), Y_{i2}(1, 0), Y_{i2}(0, 1), Y_{i2}(1, 1)) 
 \end{eqnarray*}
 We have
 \begin{eqnarray*}
 	\lefteqn{Pr(W_{i1} \mid \bX_i, Y_{i1}(0), Y_{i1}(1), Y_{i2}(0, 0), Y_{i2}(1, 0), Y_{i2}(0, 1), Y_{i2}(1, 1))=}\\&&
 	\sum_{w_2=0,1}	Pr(W_{i1}, W_{i2}=w_2 \mid \bX_i, Y_{i1}(0), Y_{i1}(1), Y_{i2}(0, 0), Y_{i2}(1, 0), Y_{i2}(0, 1), Y_{i2}(1, 1))
 	= \\&&
 	\sum_{w_2=0,1}  Pr(W_{i1} \mid \bX_i) \times Pr(W_{i2}=w_2 \mid \bX_i, W_{i1}, Y_{i1}(0), Y_{i1}(1))
 	=\\&&
 	Pr(W_{i1} \mid \bX_i) \times  \sum_{w_2=0,1} Pr(W_{i2}=w_2 \mid \bX_i, W_{i1},Y_{i1}(0), Y_{i1}(1))
 	=  Pr(W_{i1} \mid \bX_i) 
 \end{eqnarray*}
 where the first equality follows from the law of total probability and  the second equality follows from Assumption \ref{ass:LSI_AM}. Therefore Assumption \ref{ass:LSI_AM} implies Equation \eqref{eq:LSI1}. Moreover 
 \begin{eqnarray*}
 	\lefteqn{Pr(W_{i2} \mid \bX_i, W_{i1},  Y_{i1}(0), Y_{i1}(1), Y_{i2}(0, 0), Y_{i2}(1, 0), Y_{i2}(0, 1), Y_{i2}(1, 1)) =}\\&&
 	\dfrac{Pr(W_{i1}, W_{i2} \mid \bX_i, Y_{i1}(0), Y_{i1}(1), Y_{i2}(0, 0), Y_{i2}(1, 0), Y_{i2}(0, 1), Y_{i2}(1, 1))}{Pr(W_{i1} \mid \bX_i, Y_{i1}(0), Y_{i1}(1), Y_{i2}(0, 0), Y_{i2}(1, 0), Y_{i2}(0, 1), Y_{i2}(1, 1))}
 	= \\&&
 	\dfrac{Pr(W_{i1} \mid \bX_i) \times  P(W_{i2} \mid \bX_i,W_{i1}, Y_{i1}(0), Y_{i1}(1))}{ Pr(W_{i1} \mid \bX_i)}=
 	P(W_{i2} \mid \bX_i,W_{i1}, Y_{i1}(0), Y_{i1}(1)) \\
 \end{eqnarray*}
 where the first equality holds by definition, and the second equality follows from Assumption \ref{ass:LSI_AM}.
 % and the previous   result: $Pr(W_{i1} \mid \bX_i, Y_{i1}(0), Y_{i1}(1), Y_{i2}(0, 0), Y_{i2}(1, 0), Y_{i2}(0, 1), Y_{i2}(1, 1))=  Pr(W_{i1} \mid \bX_i) $.
 
 \underline{Vice-versa suppose that Equations \eqref{eq:LSI1} and \eqref{eq:LSI2} hold.} Then
 \begin{eqnarray*}
 	\lefteqn{Pr(\bW_i \mid \bX_i , Y_{i1}(0), Y_{i1}(1), Y_{i2}(0, 0), Y_{i2}(1, 0), Y_{i2}(0, 1), Y_{i2}(1, 1))=}\\&&
 	Pr(W_{i1} \mid \bX_i , Y_{i1}(0), Y_{i1}(1), Y_{i2}(0, 0), Y_{i2}(1, 0), Y_{i2}(0, 1), Y_{i2}(1, 1))
 	\times \\&&
 	Pr(W_{i2} \mid \bX_i, W_{i1},  Y_{i1}(0), Y_{i1}(1), Y_{i2}(0, 0), Y_{i2}(1, 0), Y_{i2}(0, 1), Y_{i2}(1, 1))=\\&&
 	Pr(W_{i1} \mid \bX_i)  \times  Pr(W_{i2} \mid W_{i1},   Y_{i1}(0), Y_{i1}(1), \bX_i)
 \end{eqnarray*}
 where the first equality holds by definition and the second equality follows from Equations \eqref{eq:LSI1} and \eqref{eq:LSI2}.
 
 \begin{description}
 	\item[\textit{$3.$}] \textit{Assumption \ref{ass:SI_AM} implies 
 		Assumption \ref{ass:LSI_AM}}
 \end{description}
Equation \eqref{eq:SIL1} coincides with Equation \eqref{eq:LSI1}, and Equation \eqref{eq:SIL1} implies
Equation \eqref{eq:LSI1}. Therefore Equations \eqref{eq:SIL1} and \eqref{eq:SIL2} implies 
Equations \eqref{eq:LSI1} and \eqref{eq:LSI2}. Because Assumption \ref{ass:SI_AM} is equivalent to Equations \eqref{eq:SIL1} and \eqref{eq:SIL2} and Assumption \ref{ass:LSI_AM} is equivalent to Equations \eqref{eq:LSI1} and \eqref{eq:LSI2}, we also have that Assumption \ref{ass:SI_AM} implies Assumption \ref{ass:LSI_AM}.
\begin{flushright}
$\square$
\end{flushright}

\bigskip
\if0\blind
{
\section*{Acknowledgements}

Financial support for this research was provided through the grant ``Programma Futuro in Ricerca 2012 -- RBFR12SHVV\_003'', financed by the Italian Ministero dell'Istruzione, dell'Universit\`{a} e della Ricerca.
} \fi

\if1\blind
{
  
} \fi

\bibliographystyle{agsm}
\bibliography{causality-1}

\end{document}